\documentclass[amssymb,nobibnotes,superscriptaddress,notitlepage,twocolumn]{revtex4-2}

\usepackage{enumerate,amsmath}
\usepackage{graphicx}
\usepackage{color} 

\usepackage{braket}
\usepackage{qcircuit}
\usepackage{algpseudocode}
\usepackage{algorithm}
\usepackage{ulem}

\usepackage{hyperref}
\hypersetup{
    colorlinks=true,
    linkcolor=blue,
    filecolor=magenta,   
    citecolor=magenta,
    urlcolor=cyan,
    }
\usepackage{url}

\newtheorem{thm}{Theorem}

\begin{document}

\title{Compilation of Trotter-Based Time Evolution for Partially Fault-Tolerant Quantum Computing Architecture}%

\author{Yutaro Akahoshi}
\email{akahoshi.yutaro@fujitsu.com}
\author{Riki Toshio}
\author{Jun Fujisaki}
\author{Hirotaka Oshima}
\author{Shintaro Sato}
\affiliation{Quantum Laboratory, Fujitsu Research, Fujitsu Limited, \\4-1-1 Kawasaki, Kanagawa 211-8588, Japan}
\affiliation{Fujitsu Quantum Computing Joint Research Division, Center for Quantum Information and Quantum Biology, Osaka University, \\1-2 Machikaneyama, Toyonaka, Osaka, 565-8531, Japan}

\author{Keisuke Fujii}
\affiliation{Fujitsu Quantum Computing Joint Research Division, Center for Quantum Information and Quantum Biology, Osaka University, \\1-2 Machikaneyama, Toyonaka, Osaka, 565-8531, Japan}
\affiliation{Graduate School of Engineering Science, Osaka University, \\1-3 Machikaneyama, Toyonaka, Osaka, 560-8531, Japan}
\affiliation{Center for Quantum Information and Quantum Biology, Osaka University, 560-0043, Japan}
\affiliation{RIKEN Center for Quantum Computing (RQC), Wako Saitama 351-0198, Japan}

\begin{abstract}
Achieving practical quantum speedup with limited resources is a crucial challenge in both academic and industrial communities. 
To address this, a partially fault-tolerant quantum computing architecture called ``space-time efficient analog rotation quantum computing architecture (STAR architecture)'' has been recently proposed. 
This architecture focuses on minimizing resource requirements while maximizing the precision of non-Clifford gates, essential for universal quantum computation.
However, non-deterministic processes such as the repeat-until-success (RUS) protocol and state injection  can introduce significant computational overhead. 
Therefore, optimizing the logical circuit to minimize this overhead by using efficient fault-tolerant operations is essential.
This paper presents an efficient method for simulating the time evolution of the 2D Hubbard model Hamiltonian, a promising application of the STAR architecture.
We present two techniques, parallel injection protocol and adaptive injection region updating, to reduce unnecessary time overhead specific to our architecture.
By integrating these with the existing fSWAP technique, we develop an efficient Trotter-based time evolution operation for the 2D Hubbard model.
Our analysis reveals an acceleration of over 10 times compared to naive serial compilation.
This optimized compilation enables us to estimate the computational resources required for quantum phase estimation of the 2D Hubbard model. For devices with a physical error rate of $p_{\rm phys} = 10^{-4}$, we estimate that approximately $6.5 \times 10^4$ physical qubits are required to achieve faster ground state energy estimation of the $8\times8$ Hubbard model compared to classical computation.
\end{abstract}

\maketitle
\section{Introduction} \label{sec:intro} 

Quantum computers are expected to bring exponential speedup over classical computers for various tasks, such as prime factoring~\cite{shor1999polynomial}, simulation of quantum many-body systems~\cite{abrams1999quantum,aspuru2005simulated}, and performing linear algebraic operations~\cite{harrow2009quantum}. 
Recent advancements have led to the development of so-called noisy intermediate scale quantum (NISQ) devices~\cite{preskill2018quantum}, which feature hundreds of qubits. 
Despite this progress, extracting practical quantum advantages from such NISQ devices remains challenging owing to the existence of quantum noise.

The ultimate solution to quantum noise in a scalable manner is fault-tolerant quantum computing (FTQC) using the quantum error correction (QEC) codes. 
One promising FTQC architecture is based on the surface code~\cite{KITAEV20032,10.1063/1.1499754,PhysRevLett.98.190504,PhysRevA.86.032324,Horsman_2012}, which boasts a high noise threshold and good compatibility with the superconducting qubit devices. 
These advantages have spurred several small-scale surface code experiments~\cite{zhao2022realization,krinner2022realizing,Acharya2023}. 
However, resource estimations for several tasks indicate that surface code-based FTQC architecture requires 
around $10^6$ physical qubits
to achieve meaningful quantum advantage~\cite{gidney2021factor,yoshioka2022hunting,reiher2017elucidating,goings2022reliably}. 
This is several orders of magnitude more than what current NISQ devices, with hundreds of physical qubits, can offer. 
To bridge this gap, researchers are exploring ways to reduce the resource overhead necessary for the practical quantum advantage. 
Efforts include developing new algorithms with lower resource requirements~\cite{PRXQuantum.3.010318,PhysRevLett.129.030503,https://doi.org/10.48550/arxiv.2209.11322,PRXQuantum.4.020331}, 
trading off QEC overhead for other form (such as a sampling overhead) through mitigation techniques~\cite{suzuki2022quantum,piveteau2021error}, 
and developing new QEC codes that have smaller resource overhead than the surface code~\cite{PRXQuantum.2.040101,Panteleev2021degeneratequantum,gidney2023yoked,hong2024longrangeenhanced,Bravyi_2024,goto2024manyhypercube,yoshida2024concatenate}. 

Under these circumstances, we previously introduced the
``space-time efficient analog rotation quantum computing architecture (STAR architecture)''~\cite{PRXQuantum.5.010337}, designed for the early stages of fault-tolerant quantum computing (early-FTQC).  
This architecture integrates fault-tolerant Clifford gates, protected by QEC and performed via surface code lattice surgery~\cite{Horsman_2012,Litinski2019gameofsurfacecodes}, with noisy analog rotation gates performed directly using a gate teleportation circuit consuming an ancilla state. 
By avoiding the magic state distillation~\cite{bravyi2012magic} and the Solovay--Kitaev decomposition~\cite{ross2016optimal}, the STAR architecture reduces space-time overhead of the non-Clifford gates. 
As a drawback, the ancilla state injection for the analog rotation gate is not fault-tolerant, 
thus, we need to remove the errors occurring during the state injection as much as possible. 
Recently, we 
developed a novel ancilla state injection protocol~\cite{Toshio2024}, inspired by Choi {\it et al.}~\cite{choi2023fault}. 
Although the rotation angle is restricted to be small enough, the new protocol improves the worst-case logical error rate from the original protocol of the STAR architecture~\cite{PRXQuantum.5.010337}, whose behavior is $\propto p_{\rm phys}$, to $\propto \theta p_{\rm phys}$, where $\theta$ is a small rotation angle. 
Therefore, for small angles, the logical error rate improves by a factor of $\theta$. 
In principle, this new protocol achieves sufficient precision to handle some important tasks in the early-FTQC era, such as quantum simulations of condensed-matter physics. 

For the practical application of the STAR architecture, the appropriate compilation of the input logical quantum circuit is mandatory. 
In the early-FTQC era, where resources are limited, reducing unnecessary runtime overhead is significant in achieving quantum acceleration. 
There have been several discussions on circuit compilation in the context of the surface code-based FTQC architecture~\cite{Herr_2017,Lao_2019,Lee_2022,PRXQuantum.3.020342,hamada2024efficient,tan2024sat}, where the main overhead comes from the lattice surgery operations that need additional ancilla logical qubits. 
If the number of available ancilla logical qubits is restricted, 
waiting for them to become available creates unnecessary overhead. 
Optimizing lattice surgery operations is generally NP-hard~\cite{Herr_2017}, but several heuristic optimization techniques have been proposed~\cite{Lao_2019,Lee_2022,PRXQuantum.3.020342,hamada2024efficient,tan2024sat}.
In the STAR architecture, the situation differs owing to the use of the direct analog rotation gates. 
A major advantage of STAR architecture is that analog rotation gates can easily be performed in parallel without any space overhead, thanks to the absence of distillation blocks. 
Compilers should utilize this advantage to reduce runtime. 
On the other hand, the STAR architecture's runtime overhead also depends on the repeat-until-success (RUS) protocol and the ancilla state injection protocol. 
The non-deterministic nature of the RUS protocol complicates pre-compiling its execution order. 
If ancilla state injections rarely succeed, the runtime per each RUS trial increases, and the success rate of state injections decreases with longer RUS protocol durations owing to increasing rotation angles. 
Therefore, to maximize the benefits of STAR architecture, a ``real-time'' strategy is needed to minimize the runtime overhead of the rotation gates, thereby suppressing the worst-case overhead of long-lasting RUS protocols. 

In this paper, we present a method to compile a representative task while reducing the overhead of the RUS protocols. 
We focus on the quantum simulation of the 2D Hubbard model~\cite{doi:10.1098/rspa.1963.0204}, expected to be the benchmark of the practical quantum advantage in the near future~\cite{Campbell_2022, Kivlichan2020improvedfault,yoshioka2022hunting}. 
To utilize the small-angle rotation gate in STAR architecture, we employ the Trotterization approach for the quantum simulations. 
We reduce the time overhead of the RUS protocol by employing the fSWAP network~\cite{PhysRevLett.120.110501}, which localizes the Hamiltonian's interaction terms, allowing parallel execution of these local rotation gates.
Additionally, we introduce two ideas to decrease the ancilla state injection overhead:
parallel injection and adaptive injection region updating. 
Numerical simulations of the RUS protocol using these techniques indicate an over 80 \% reduction in time overhead when the injection success rate is low. 
Combining all these technical elements, we derive a formula for the execution time of the time evolution operator, which can be applied to any estimation of any computational task using Trotter-based time evolution of the 2D Hubbard model. 
To demonstrate the application of the formula, we estimate the total runtime and required number of physical qubits for the quantum phase estimation (QPE) algorithm designed for the early-FTQC era~\cite{PRXQuantum.4.020331}. 
Using parameters from previous studies, we estimate that for a physical error rate of $p_{\rm phys} = 10^{-4}$, an $8 \times 8$ Hubbard model simulation can be performed in about 10 hours with $6.5 \times 10^4$ physical qubits. 
This is approximately $10^3$ faster than the classical algorithm, given in Ref.~\cite{yoshioka2022hunting}, 
suggesting that the STAR architecture could offer practical quantum advantages with a smaller quantum device than previously expected. 

This paper is organized as follows. 
Sec.\ref{sec:model} provides a brief introduction to the STAR architecture, which forms the foundation of our discussion. 
Sec.\ref{sec:compile}, the main contribution of this paper, explores 
an efficient operation schedule of the Trotter-based time evolution operator within the STAR architecture. 
We start by introducing the Trotterization of the time evolution operator and the Hamiltonians considered in this study. 
Key elements of our compilation, namely the fSWAP and the parallel RUS execution, are discussed in detail. 
We then demonstrate 
how to estimate the total execution time of a single Trotter step, measured in units of the lattice surgery operation clock. 
Sec. \ref{sec:phaseest} applies our compilation results to the QPE algorithm.
We estimate the execution time and the number of physical qubits based on our findings and the proposed algorithm~\cite{PRXQuantum.4.020331}. 
Sec.\ref{sec:summary} summarizes our study and briefly discuss some issues to be addressed in future studies. 

\section{Overview of STAR architecture} \label{sec:model}
Before delving into the detailed discussion on compilation, 
it is essential to outline how we perform universal quantum computing. 
Given our unique approach, we first review its features for readers who may be unfamiliar with it.

In this paper, we employ the STAR architecture~\cite{PRXQuantum.5.010337} as the foundation of universal quantum computing. 
The STAR architecture is originally proposed in Ref.~\cite{PRXQuantum.5.010337} aiming to bridge the gap between NISQ and FTQC. 
The key features of this architecture are the following: 
\begin{enumerate}[(i)]
    \item Fault-tolerant Clifford gates using QEC.
    \item Direct analog rotation gates with reasonably clean ancilla states. 
    \item Mitigating remnant noise in the analog rotation gate by quantum error mitigation (QEM) techniques (e.g., probabilistic error cancellation (PEC)).
\end{enumerate}
By performing the analog rotation gates directly, we can avoid the need for magic state distillation~\cite{bravyi2012magic} and Solovay--Kitaev decomposition~\cite{ross2016optimal}. 
This significantly reduces the number of physical qubits and execution time of non-Clifford gates. 

\begin{figure}
    \centering
        \mbox{
        \Qcircuit @C=1em @R=.7em {
        \lstick{\ket{\psi}} & \multigate{1}{M_{ZZ}} & \gate{Z}        & \qw & \\
        \lstick{\ket{m_{\theta}}}    & \ghost{M_{ZZ}}        & \gate{M_X} \cwx &
        }
        }
    \caption{
    Quantum circuit to implement $R_{Z}(\theta)$. $M_{ZZ}$ and $M_X$ represent the measurements of $ZZ$ and $X$, respectively. 
    Based on the measurement value $m_{ZZ}$ of $M_{ZZ}$, the output state is either $R_{Z}(\theta) \ket{\psi}$ ($m_{ZZ} = +1$) or $R_{Z}(-\theta) \ket{\psi}$ ($m_{ZZ} = -1$).
    The circuit is repeated until obtaining the desired state $R_{Z}(\theta) \ket{\psi}$. }
    \label{fig:rotation_circuit}
\end{figure}
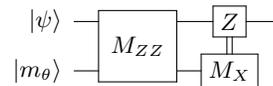

In this study, we employ the rotated planar surface code~\cite{Horsman_2012} as a logical qubit, also referred to as a ``patch'', and use lattice surgery~\cite{Horsman_2012,Litinski2019gameofsurfacecodes} to implement logical Clifford gates. 
Detailed descriptions of the logical Clifford operation used in this study are summarized in Appendix.\ref{appx:qecandls}. 
Regarding the direct analog rotation gate, 
we employ the RUS approach~\cite{Cody_Jones_2012,reiher2017elucidating,PRXQuantum.5.010337} using the circuit in Fig.~\ref{fig:rotation_circuit}. 
In general, the injection of the ancilla state, $\ket{m_\theta} \equiv R_Z(\theta) \ket + = \frac{1}{\sqrt{2}}(e^{-i \theta} \ket 0 + e^{+i \theta} \ket 1)$, cannot be fault-tolerant owing to the analog rotation angle. 
To reduce noise, we employ a recently proposed ancilla state injection protocol~\cite{Toshio2024}, We briefly explain the RUS protocol and the state injection protocol in Appendix~\ref{appx:rotationgate}. 
This protocol, although restricted to small rotation angles, drastically improves the logical error rate of the ancilla state. 
Since the Trotter-based time evolution operator is a product of small angle rotation gates, this protocol is highly suitable for our purposes. 
As discussed in Ref.~\cite{Toshio2024}, by assuming the circuit-level noise model with error probability $p_{\rm phys}$, the noise channel of the direct small-angle rotation gate can be reduced to a probabilistic $Z$ error channel, whose worst-case logical error rate is given as, 
\begin{equation} \label{eq:ver2_wcler}
    \epsilon_{{\rm RUS}, \theta^*} \approx \alpha_{\rm RUS} \theta^{*} p_{\rm phys}, 
\end{equation}
where $\alpha_{\rm RUS}$ is a factor accounts for noise accumulation during RUS trials. 
$\alpha_{\rm RUS}$ depends not only on the rotation angle $\theta^*$ but also on the parameters chosen for the injection protocol. 
Numerical simulations~\cite{Toshio2024} suggest that
$\alpha_{\rm RUS} / k \approx 0.40$ (where $k$ is a parameter of the injection protocol. See Appendix~\ref{appx:rotationgate}), and we use this value in our discussion. 
Once the noise channel is specified, we can mitigate it by QEC techniques. 
In this study, we employ the probabilistic error cancellation (PEC)~\cite{PhysRevLett.119.180509,PhysRevX.8.031027}. 
The PEC allows us to compensate the noise channel with an additional sampling overhead, 
\begin{equation} \label{eq:pec_overhead}
    \gamma^2(\theta^*) \approx e^{4 \epsilon_{{\rm RUS}, \theta^*}}. 
\end{equation}
In Sec.~\ref{sec:phaseest}, we use this equation to estimate the actual runtime of the quantum algorithm including the overhead intruduced by the PEC. 

\section{Compilation of the Trotter-based time evolution of 2D Hubbard model for STAR architecture} \label{sec:compile}
In this section, we discuss the compilation of the Trotter-based time evolution of the 2D Hubbard model, tailored for the STAR architecture.
We begin by introducing the Trotter decomposition of the time evolution operator and defining the 2D Hubbard model Hamiltonian. 
Next, we outline our compilation strategy, emphasizing the core techniques used.
After that, we introduce core techniques in our compilation, 
These include the insertion of fermionic SWAP (fSWAP) network and the execution of the parallel RUS protocol. 
Finally, we sum up all these elements to determine the execution time of the single Trotter step. 
We also compare our compilation results with a naive sequential compilation approach. 

\subsection{Trotter-based time evolution operator}
Consider the Hamiltonian given as 
\begin{equation} \label{eq:Hamiltonian_paulidecomp}
    H = \sum_{j=0}^{L-1} c_j P_j, 
\end{equation}
where $P_j$ represents a multi-qubit Pauli operator, $c_j$ denotes a real coefficient, and 
$L$ is the total number of terms. 
We aim to perform the time evolution operator $e^{-iH\tau}$, 
but directly exponentiating the Hamiltonian is impractical owing to its exponentially large dimension. 

To simplify this, we use the Trotter decomposition, which approximates the time evolution operator with easy-to-perform steps, introducing a small error. 
We divide the time interval $\tau$ into $r$ steps and define $\Delta \tau = \tau / r$. 
The second-order Trotter decomposition is given as, 
\begin{equation}
    e^{-i H \Delta \tau} \approx \prod_{j=0}^{L-1} e^{-i c_j P_j \Delta \tau / 2} \prod^{0}_{j=L-1} e^{-i c_j P_j \Delta \tau / 2} \equiv e^{-i H_{\rm eff} \Delta \tau},
\end{equation}
where $\Delta \tau$ is sufficiently small. 
The approximation error of the second-order Trotterization behaves as~\cite{Kivlichan2020improvedfault},
\begin{equation}
    || e^{-i H \Delta \tau} - e^{-i H_{\rm eff} \Delta \tau} || \leq W \Delta \tau^3, 
\end{equation}
where $W$ is a constant called ``Trotter error norm''~\cite{Kivlichan2020improvedfault}. 
The error of the $n$-th energy eigenvalue $E_{{\rm eff},n}$ calculated by $e^{-i H_{\rm eff} \Delta \tau}$ is also bounded as~\cite{Kivlichan2020improvedfault} 
\begin{equation} \label{eq:2nd_trott_err}
    |E_n - E_{{\rm eff}, n}| \leq W \Delta \tau^2. 
\end{equation}
The time evolution of $e^{-iH\tau}$ can be approximated by the sequence of the time evolution of $e^{-iH \Delta \tau}$, 
\begin{equation} \label{eq:2nd_trott_formula}
    e^{-iH\tau} = \left( \prod_{j=0}^{L-1} e^{-i c_j P_j \Delta \tau / 2} \prod^{0}_{j=L-1} e^{-i c_j P_j \Delta \tau / 2} \right)^r + {\mathcal O}(1/r^2)
\end{equation}
As shown, the Trotter-based time evolution is a product of small-angle Pauli rotations. 
In Sec.~\ref{sec:phaseest}, Eq. (\ref{eq:2nd_trott_err}) is used to determine the division number $r$ for the Trotter decomposition via $\Delta \tau = \tau / r$. 

\subsection{Hamiltonian}
In this study, we consider the 2D Hubbard model Hamiltonian on an $N \times N$ square lattice. 
The 2D Hubbard model is a simple model that captures the electronic and magnetic behavior in solid materials~\cite{RevModPhys.66.763,annurev-conmatphys-031620-102024}.
The Hamiltonian of the 2D Hubbard model is defined as, 
\begin{equation}
    H = -t \sum_{\langle i,j \rangle, \sigma} (c^{\dag}_{i, \sigma}c_{j, \sigma} + {\rm h.c.}) + U \sum_{i} n_{i, \uparrow} n_{i, \downarrow}, 
\end{equation}
where $c^{(\dag)}_{i, \sigma} \ (\sigma = \uparrow, \downarrow)$ represents the creation (annihilation) operator of the fermion 
and $n_{i, \sigma} = c^{\dag}_{i, \sigma} c_{i, \sigma} \ (\sigma = \uparrow, \downarrow)$ denotes the number operator of up and down spin component on the site $i$. 
The notation $\langle i,j \rangle$ indicates pairs of adjacent sites on the 2D lattice, while 
$t$ and $U$ are parameters indicating the hopping strength and repulsive potential strength, respectively. 
The Jordan-Wigner (JW) transformation~\cite{PhysRevA.65.042323} with a certain index ordering transforms 
the Hamiltonian as follows, 
\begin{eqnarray}
  H &=& -\frac{t}{2} \sum_{\langle i,j \rangle, \sigma} (X_{i,\sigma} X_{j,\sigma} + Y_{i,\sigma} Y_{j,\sigma}) Z^{\leftrightarrow}_{i,j,\sigma} \nonumber \label{eq:2Dhubbard} \\ 
  &+& \frac{U}{4} \sum_{i} Z_{i,\uparrow} Z_{i,\downarrow} - \frac{U}{4} \sum_{i,\sigma} Z_{i,\sigma}, \\ 
  Z^{\leftrightarrow}_{i,j,\sigma} &=& \prod_{k \in P_{i,j}} Z_{k,\sigma},
\end{eqnarray}
where $P_{i,j}$ indicates a set of sites between $i$ and $j$ in the one-dimensional ordering of the JW transformation. 
The ordering we choose in this study will be discussed later. 
Originally, the Hamiltonian consists only of local interaction terms. However, the JW transformation introduces non-local multi-qubit Pauli operators to maintain the fermionic commutation relations in the one-dimensional mapping. 
As discussed in Ref.~\cite{Campbell_2022}, the last term of Eq.(\ref{eq:2Dhubbard}) is proportional to the total electron number operator up to some constant factor. 
Generally, the simulation is performed in a subspace with a fixed number of electrons, 
allowing us to omit the last term of Eq.(\ref{eq:2Dhubbard}) by absorbing the constant factor into the energy eigenvalues. 
Therefore, the Hamiltonian we consider is reduced to
\begin{eqnarray} \label{eq:2Dhubbard_final}
  H = -\frac{t}{2} \sum_{\langle i,j \rangle, \sigma} (X_{i,\sigma} X_{j,\sigma} + Y_{i,\sigma} Y_{j,\sigma}) Z^{\leftrightarrow}_{i,j,\sigma} + \frac{U}{4} \sum_{i} Z_{i,\uparrow} Z_{i,\downarrow} \nonumber \\
\end{eqnarray}
In this study, we apply the open boundary condition. 
The parameters $t, U$ are chosen as $t = 1, U = 4$, which consistent with previous studies~\cite{Kivlichan2020improvedfault, yoshioka2022hunting}. 

\subsection{Compile strategy} \label{sec:strategy}
In the previous section, we noted that the JW transformation results in non-local multi-qubit Pauli terms like $P = XZ\dots ZX$ in the Hamiltonian, making these interaction terms challenging to execute in parallel. 
However, to reduce the runtime overhead of the RUS protocol of rotation gates, 
it is better to perform these gates in parallel. 
Unlike the existing FTQC architecture, the STAR architecture can dynamically prepare the ancilla states as needed without any ancilla state factory. 
This capability allows for parallel execution of rotation gates without increasing qubit overhead.
According to this observation, we consider the following compile strategy in this study:
\begin{itemize}
    \item Localizing as many interaction terms as possible in terms of the Pauli operator,
    \item Performing multiple local interaction terms in parallel using small-angle rotation gates,
    \item Optimizing runtime overhead by fine-tuning patch arrangements and lattice surgery operation schedules. 
\end{itemize}
We achieve the first objective 
by inserting fermionic SWAP operations~\cite{PhysRevLett.120.110501}, which rearrange the JW transformation assignments to make subsets of interaction terms local. 
Using the STAR architecture's rotation gates, we then perform multiple local interaction terms in parallel. 
Proper patch arrangements are important to maintain the parallelism and avoid disruptions from the lattice surgery routing. 
Additionally, we mitigate the overhead of the state injection in the RUS protocol through parallel injection and adaptive injection region updating. 
In the following, we will delve into each technical component before consolidating these techniques to present the final compilation results.

\subsection{Fermionic SWAP insertion}
To perform quantum many-body simulations on quantum computers, 
we need to map fermions to qubit states while preserving fermionic statistics. 
The typical approach for this mapping is the JW transformation. 
However, as seen in Eq.~(\ref{eq:2Dhubbard}), this transformation generally disrupts the locality of interaction terms.
The fermionic SWAP (fSWAP) operation~\cite{PhysRevLett.120.110501} offers a solution by enabling all interaction terms in the Hamiltonian to be performed locally at the cost of additional operations. 
The basic idea of this method is simple: changing the JW transformation mapping to position all interacting pairs of spin orbits adjacently. 
The fSWAP operator acting on the pair of spin orbits $p, q$ is defined as follows, 
\begin{equation} \label{eq:fswapdef}
    {\rm fSWAP}_{p,q} = 1 + c^{\dag}_p c_q + c^{\dag}_q c_p - c^{\dag}_p c_p - c^{\dag}_q c_q.
\end{equation}
This operator acts on the fermionic creation and annihilation operators as 
\begin{eqnarray}
    {\rm fSWAP}_{p,q} c^{\dag}_p {\rm fSWAP}^{\dag}_{p,q} &=& c^{\dag}_q, \\
    {\rm fSWAP}_{p,q} c_p {\rm fSWAP}^{\dag}_{p,q} &=& c_q,
\end{eqnarray}
Therefore, the fSWAP operator interchanges two spin orbits while keeping Fermi statistics. 
In this study, we only apply the adjacent fSWAP operation, namely $q = p + 1$. 
Using the JW transformation, we obtain a simple matrix form of the adjacent fSWAP operator, 
\begin{equation}
    {\rm fSWAP}_{p, p+1} = 
    \begin{pmatrix}
        1 & 0 & 0 & 0 \\
        0 & 0 & 1 & 0 \\
        0 & 1 & 0 & 0 \\
        0 & 0 & 0 & -1 
    \end{pmatrix},
\end{equation}
which is just a sequence of the standard SWAP gate and the CZ gate. 

\begin{figure*}[tbp] 
    \centering
    \includegraphics[width=80mm, clip]{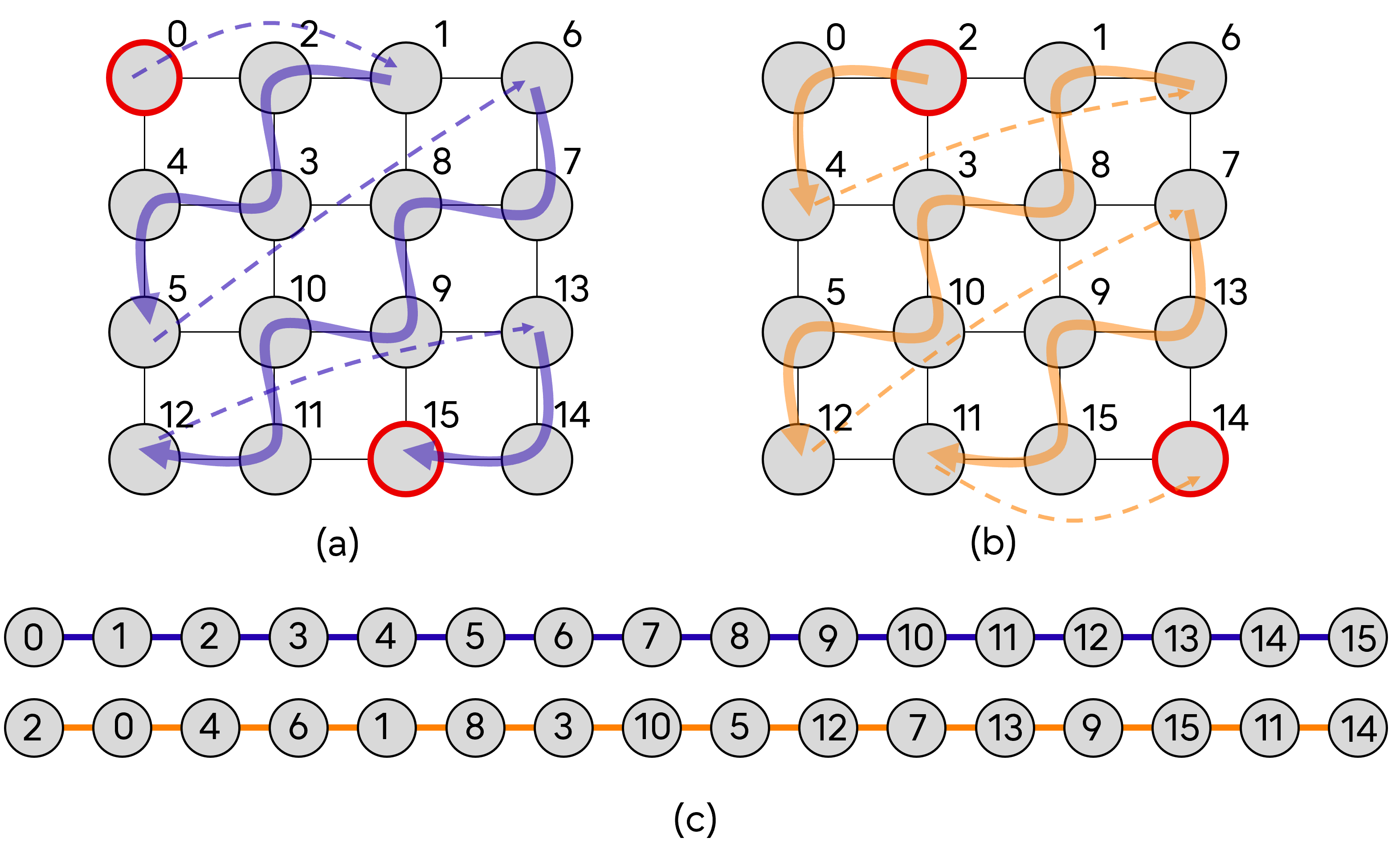} \hspace{10mm}
    \includegraphics[width=80mm, clip]{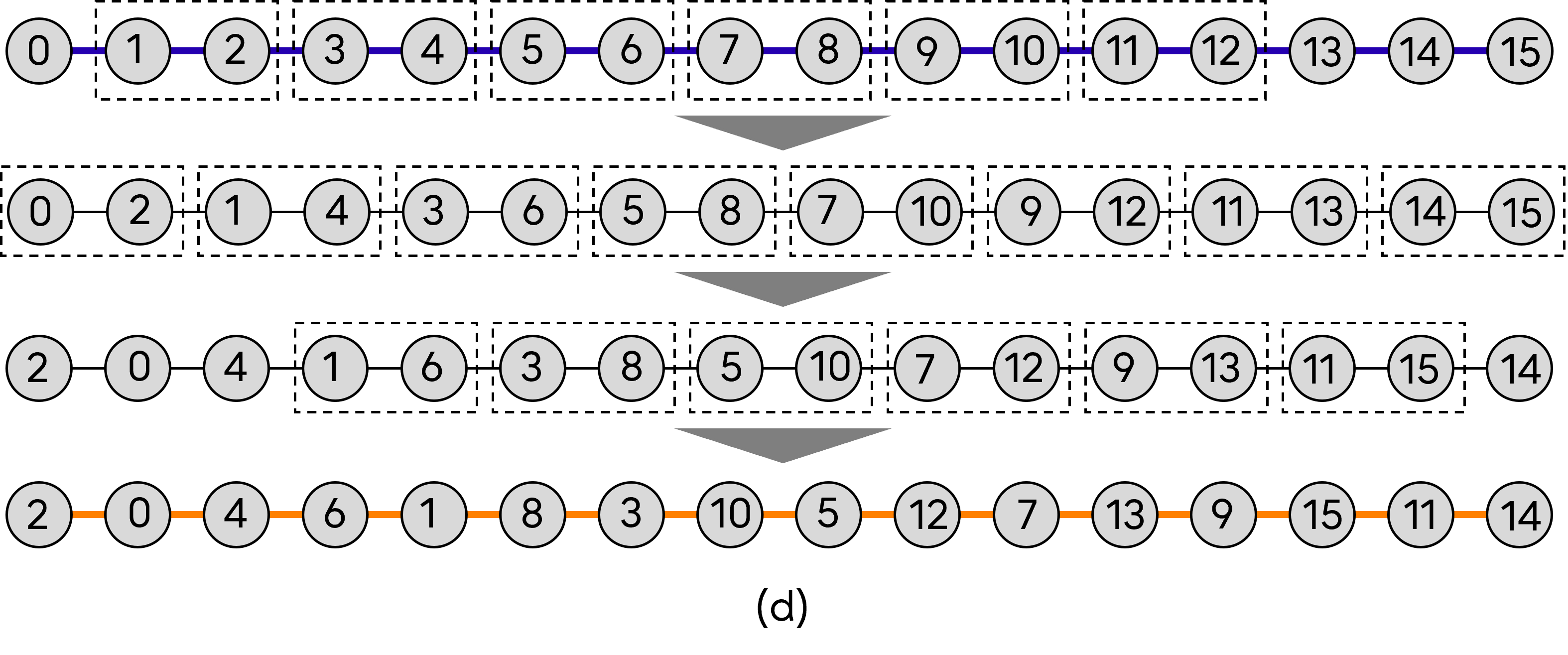}
    \caption{
    Summary of fSWAP strategy used in this study, demonstrated by a $4 \times 4$ Hubbard model. 
    (a) and (b) show the JW ordering considered in this study. 
    We show only the up-spin components here, but the same orderings are also used in down-spin components. 
    The beginning and end of the ordering are marked by red circles. 
    The union of these two orderings covers all hopping interaction edges. 
    (c) One-dimensional representation of the JW ordering from (a) and (b), with indices inserted from (a) to visualize the movement of each qubit via the fSWAP operations. 
    (d) Adjacent fSWAP operations required to change the JW ordering from (a) to (b), with each layer performing fSWAP operations on pairs of qubits connected by dashed lines. It takes three layers of fSWAP operations in this case. 
    }
    \label{fig:fswap4x4}
\end{figure*}

The key challenges lies in inserting the fSWAP gates to achieve fully local interactions while minimizing time overhead (circuit depth).
Previous studies have discussed this issue, and in this study, we employ the strategy in Ref.~\cite{hagge2022optimal}. 
We illustrate this with a $4 \times 4$ Hubbard model, as depicted in Fig.~\ref{fig:fswap4x4}. 
We consider two different qubit orderings of the JW transformations, the union of them covers all interaction terms (ordering (a) and (b) in Fig.~\ref{fig:fswap4x4}). 
We start with ordering (a), where 
half of the hopping interactions can be performed locally since they are already adjacent. 
After performing these adjacent interactions, we perform fSWAP operations to transform the current ordering into the other one (ordering (b) in Fig.~\ref{fig:fswap4x4}). 
According to Theorem 6 in Ref.~\cite{hagge2022optimal}, this transformation requires $N-1$ fSWAP layers for the $N \times N$ Hubbard model. 
In Fig.~\ref{fig:fswap4x4} (d), we show the explicit fSWAP operations needed to transform ordering (a) to (b). 
Finally, we perform the remaining half of the hopping interactions are in the same manner as in ordering (a). 
The details of the lattice surgery implementation will be discussed in Sec.~\ref{sec:compile_ls}.  

\subsection{Parallel RUS execution}
Once hopping interactions become local, the STAR architecture can easily execute them in parallel. 
In this section, 
we will discuss the average number of RUS trials required for parallel execution and its advantages over the serial RUS execution. 
We will also cover strategies to reduce the runtime overhead of the injection protocol, assuming a basic understanding of lattice surgery (details provided in Appendix A).

\begin{figure}
    \centering
    \includegraphics[width=80mm, clip]{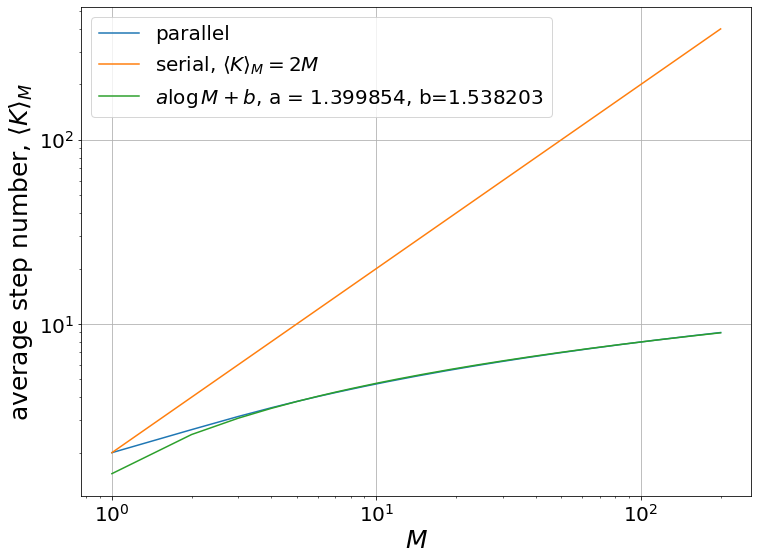}
    \caption{Average number of RUS trials for $M$ parallel RUS executions (blue line). 
    The figure also shows the average trial number of serial RUS execution and a fitting result for parallel RUS execution. 
    When $M$ is large, the scaling of the parallel RUS execution is well described by the log function. 
    }
    \label{fig:averageRUSparallel}
\end{figure}

Let us consider the case where we perform $M$ independent rotation gates in parallel.
To estimate the average number of RUS trials need to complete all $M$ processes, 
we first calculate the probability that the parallel RUS execution finishes on the $K$-th RUS trial, $P_{K}^{M}$, and then calculate the expectation value of $K$. 
As derived in Appendix~\ref{appx:averageRUSderiv}, the expectation value of $K$ behaves as, 
\begin{equation} \label{eq:aveRUS}
    \begin{array}{ll}
    \langle K \rangle_M &= \sum_K K P_K^M \\
    &= \sum_K K \left[ (1-2^{-K})^M - (1-2^{-K+1})^M \right]. 
    \end{array}
\end{equation}
One can easily verify that if $M = 1$, we can reproduce $\langle K \rangle_{M=1} = \sum_K K / 2^K = 2$. 
We plot $\langle K \rangle_M$ for different $M$ in Fig.~\ref{fig:averageRUSparallel}. 
As you can see, the average number of RUS trials behaves log-like for large $M$; 
therefore, the parallel execution is drastically faster than the serial one if the injection protocol overhead is negligible. 

\begin{figure}
    \centering
    \includegraphics[width=30mm,clip]{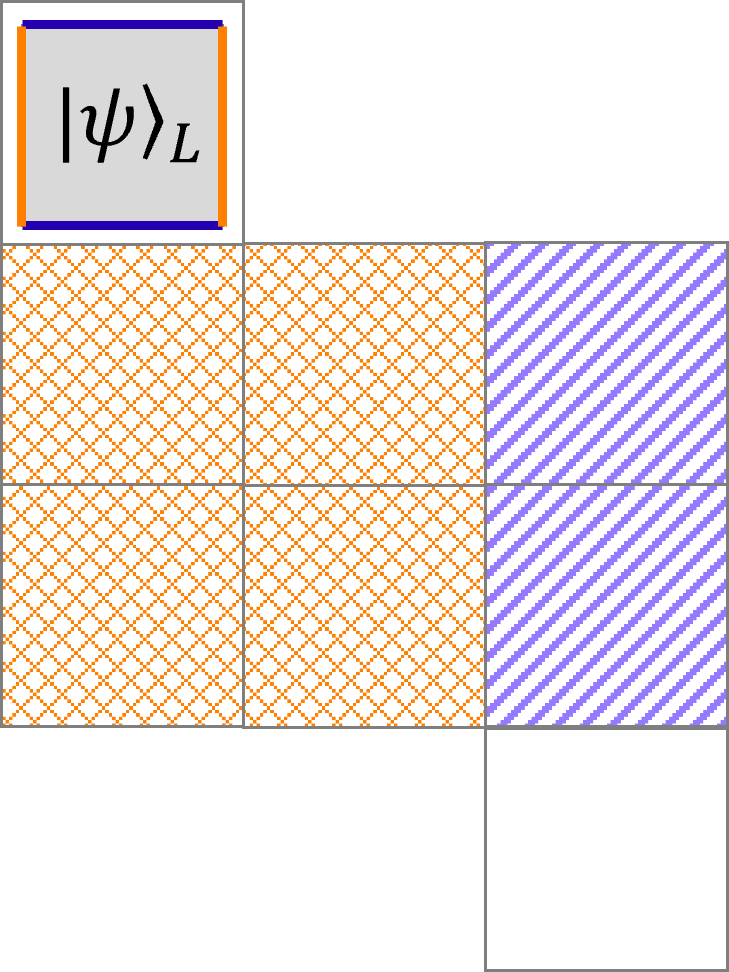}
    \includegraphics[width=45mm, clip]{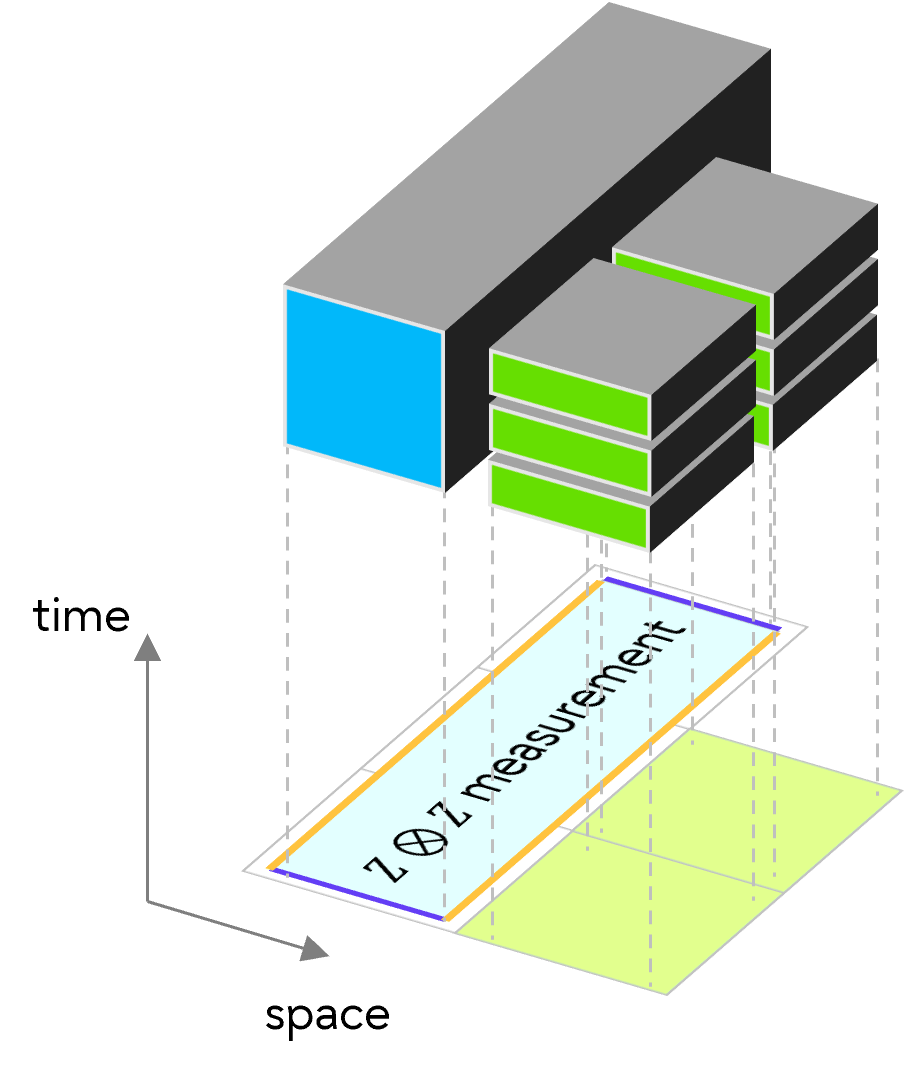}
    \caption{
    Example of parallel injection. 
    (left) Example of a free ancilla region that can be used for the parallel injection. 
    We consider acting a rotation gate on the logical patch $\ket{\psi}$, while purple-shaded patches are already used in other operations. 
    In this case, four orange-shaded patches can be used for the injection. 
    The white patch cannot be used for the injection because there is no path connecting the target logical patch with this white patch.
    (right) Parallel injection along the time direction. During the $ZZ$ measurement in the RUS circuit, we perform state injection for the next RUS trial on free patches in the injection region. Since the injection trial can be performed in a shorter time than a single lattice surgery clock, we can repeat the injection several times (in this figure, three trials). 
    If certain trials succeed, we can immediately execute the next RUS trial. 
    }
    \label{fig:parallelinjection}
\end{figure}

In practice, however, the injection overhead can be significant in some cases. 
Let us then discuss strategies to reduce the injection overhead and maximize the benefits of the parallelism. 
As discussed in Sec.~\ref{sec:model}, the injection protocol is non-deterministic and can result in large time overhead when it fails repeatedly. 
Moreover, 
longer RUS protocols lead to larger rotation angles (for the $K$-th trial, the rotation angle is $2^K$ times larger than the original), 
reducing the success rate since it depends on the rotation angle via $p_{\rm ideal}$. 
Fortunately, the injection protocol uses only one logical patch to inject an ancilla state $\ket{m_{\theta}}$. 
This allows parallel execution if there are free patches available to 
connect to the logical patch where the rotation gate is to be performed.
Figure~\ref{fig:parallelinjection} (left) shows one such example. 
We can also pre-inject the 
ancilla state needed for the next RUS trial in advance during the $ZZ$ measurement of the current RUS trial (see Fig.~\ref{fig:rotation_circuit}) using available ancilla patches. 
This ``space'' parallelism exponentially reduces the injection time overhead:
if $l$ free ancilla patches are available, the failure rate improves from $P_{\rm fail}$ to $P_{\rm fail}^l$.
Furthermore, we can heap up the injection protocol along the ``time'' direction, 
performing several trials of the injection within a single lattice surgery clock cycle (see Fig.~\ref{fig:parallelinjection} (right)). 
This increases the number of injection trials further. 
Considering 
``space'' and ``time'' parallelism strategies significantly minimizes injection overhead until rotation gate is executed. 

\begin{figure}
    \centering
    \includegraphics[width=80mm, clip]{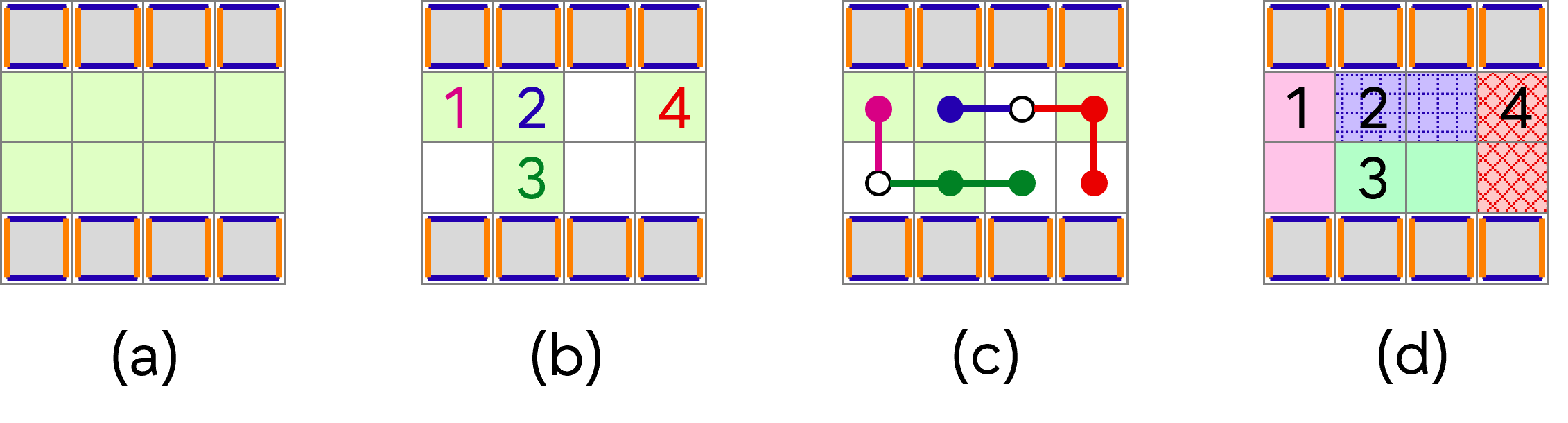}
    \caption{
    Schematic description of the updating algorithm for the injection region during parallel $Z$ rotations involving 8 patches. 
    (a) Initial injection region. Each RUS process has a single injection patch (green) adjacent to the target patch (gray). 
    (b) After several RUS trials, 4 RUS processes complete and the corresponding injection regions become free (white patches). 
    The updating algorithm then assigns these free patches to the ongoing 4 RUS processes (numbered 1 -- 4). 
    (c) To update the injection region for the ongoing processes, the algorithm expands the region clusters until the cluster can no longer grow, by adding adjacent free nodes step by step. 
    There are two types of boundaries in the cluster: isolated (filled circles) and conflicted (open circles). 
    A conflicted node means that different two region clusters add it at the same cluster-growing step. 
    (d) Final result of the update algorithm. 
    Isolated nodes can be assigned immediately to the respective injection region. 
    The conflicted nodes should be assigned to a smaller injection region (here, the size of the injection region is defined as the number of patches in the region) to balance the size of each injection region. 
    As a result, each injection region becomes size 2 in this example. 
    }
    \label{fig:updateinjectionregion}
\end{figure}

The remaining challenge is how to assign free ancilla patches to the ongoing RUS processes. 
This assignment needs to be dynamic, adjusting as new free ancilla patches become available.
To achieve this, we introduce the assignment algorithm based on a simple cluster-growing method, as illustrated in Fig~\ref{fig:updateinjectionregion}. 
This algorithm evenly distributes free patches among ongoing RUS processes. 
Thanks to this adaptive injection region updating algorithm, 
we can suppress the worst-case runtime overhead caused 
from the long-lasting RUS processes and their low success rates owing to large rotation angles. 
Since long-lasting RUS processes are rare, most RUS processes in parallel execution will have already completed, leaving many free ancilla patches available for assignment to the remaining processes. 
As a result, massively parallelized state injection compensates for its low success rate. 
Numerical simulations support this scenario, showing that combining the parallel injection with the adaptive assignment algorithm effectively reduces the runtime overhead of parallel RUS execution.


\subsection{Lattice surgery implementations} \label{sec:compile_ls}
In this section, we summarize the lattice surgery implementations needed for the Trotter-based time evolution. 
We begin by introducing a suitable logical patch arrangement and then discuss lattice surgery implementations for each component of the Trotter-based time evolution operator of the 2D Hubbard model. 
The basic operations of lattice surgery and logical Clifford gates used in this section are summarized in Appendix~\ref{appx:qecandls}. 

\begin{figure}
    \centering
    \includegraphics[width=70mm, clip]{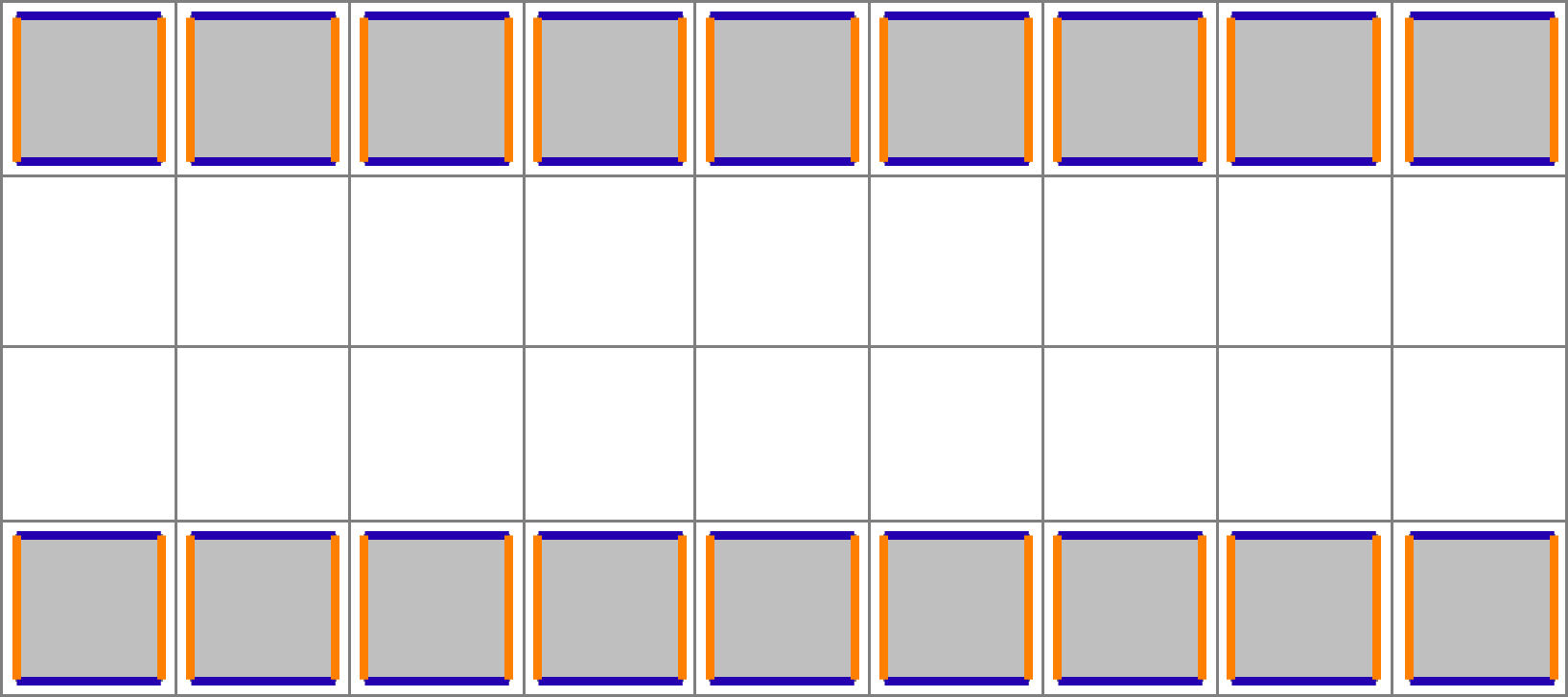}
    \caption{Patch arrangement considered in this study. 
    Gray patches indicate logical data qubits. 
    White patches in the middle are used for lattice surgery routing or the ancilla state injection. }
    \label{fig:patch2nf2f}
\end{figure}

As mentioned in Sec.~\ref{sec:strategy}, the logical patch arrangement is important to avoid unnecessary overhead from routing lattice surgery operations. 
For our purpose, the local interaction terms and the fSWAP operations are nearest-neighbor two-qubit operations. 
Therefore, a large routing region is unnecessary to perform these lattice surgery operations. 
Additionally, the most complicated part, the hopping interactions, are performed independently for each spin component, suggesting that dividing the logical patch arrangement into two regions that can operate separately is beneficial. 
Considering these requirements, we propose the logical patch arrangement shown in Fig.~\ref{fig:patch2nf2f}. 
This set-up uses $2n$ logical patches to allocate $n$ logical data qubits. 
The logical data qubits are arranged at the upper and lower sides, while the middle region is reserved for routing lattice surgery operations. 
Each logical data qubit has an adjacent ancilla patch, allowing for single-qubit rotation gates on all logical qubits simultaneously.
As discussed later, the upper-side (lower-side) logical qubits mostly carry the spin up (down) components, except for the execution of $ZZ$ rotation terms. 
\begin{figure}
    \centering
    \includegraphics[width=40mm, clip]{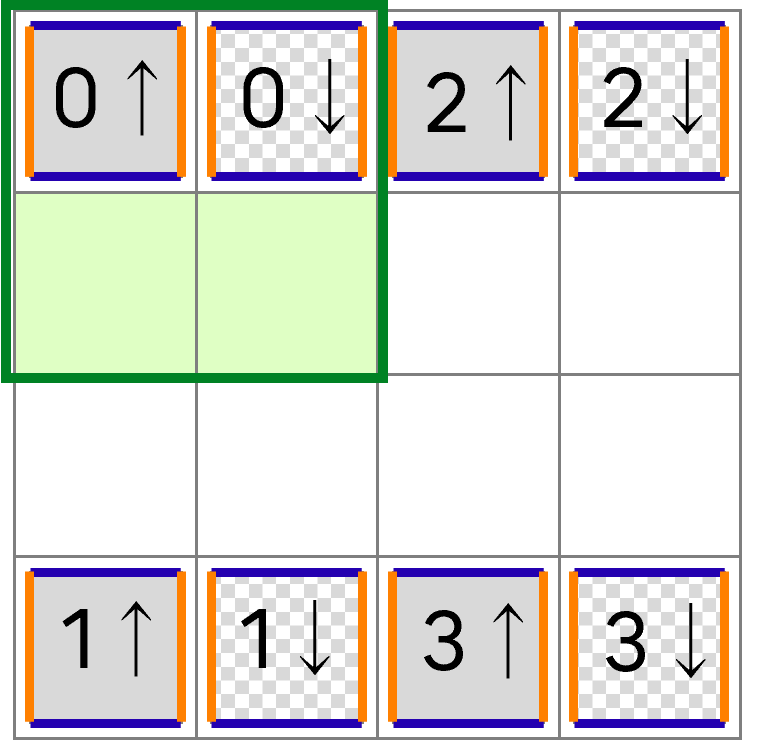}
    \caption{
    Initial assignment of logical information and execution of $ZZ$ rotation terms. 
    This figure shows only part of the whole patch arrangement here. 
    Spin-up and spin-down components are placed on adjacent patches. 
    $ZZ$ rotation terms are then executed in parallel. 
    The representative RUS process is indicated as a quoted area by a green solid line, 
    and its initial injection region consists of green-colored patches. 
    }
    \label{fig:init_Z_ZZ}
\end{figure}
\begin{figure}
    \centering
    \includegraphics[width=60mm, clip]{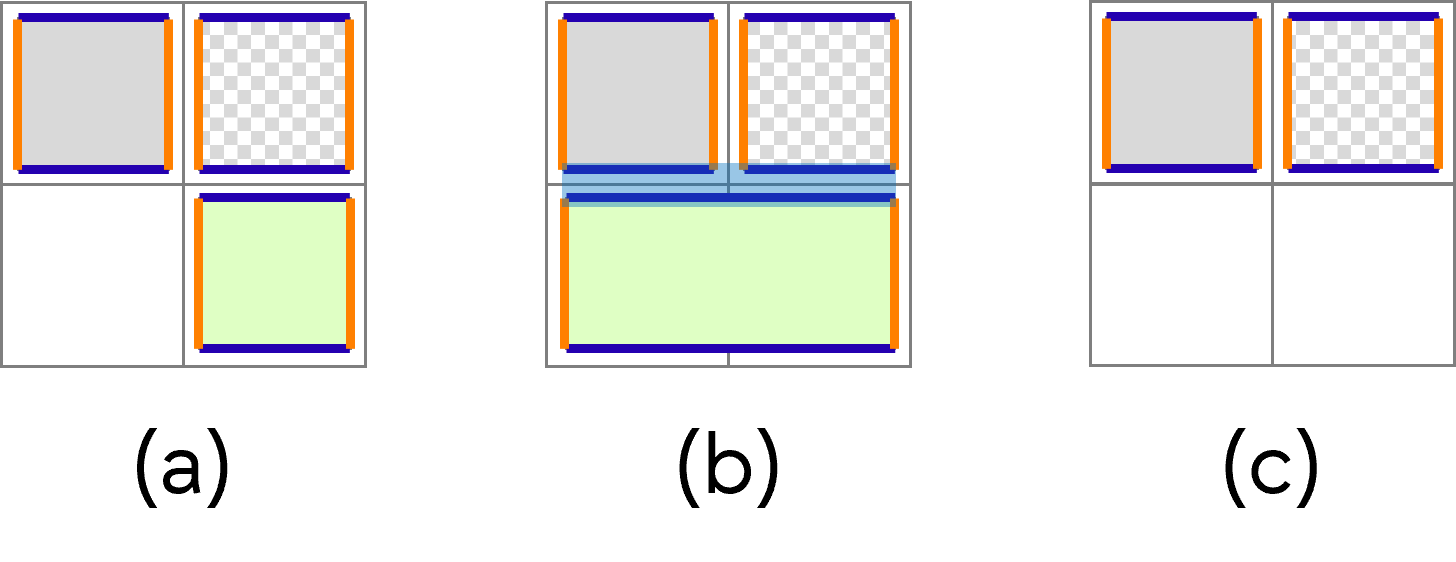}
    \caption{Lattice surgery implementation of the $ZZ$ rotation gate. 
    (a) The ancilla state $\ket{m_{\theta}}$ is prepared on a free patch (green patch). 
    (b) To measure the $ZZZ$ operator, the ancilla state is expanded to a rectangular shape and merged with the target two patches. 
    (c) After splitting, the destructive logical $X$ measurement is performed on the ancilla patch. The whole procedure takes 2 logical clocks.
    } 
    \label{fig:ZZrot_ls}
\end{figure}
\begin{figure}
    \centering
    \includegraphics[width=80mm, clip]{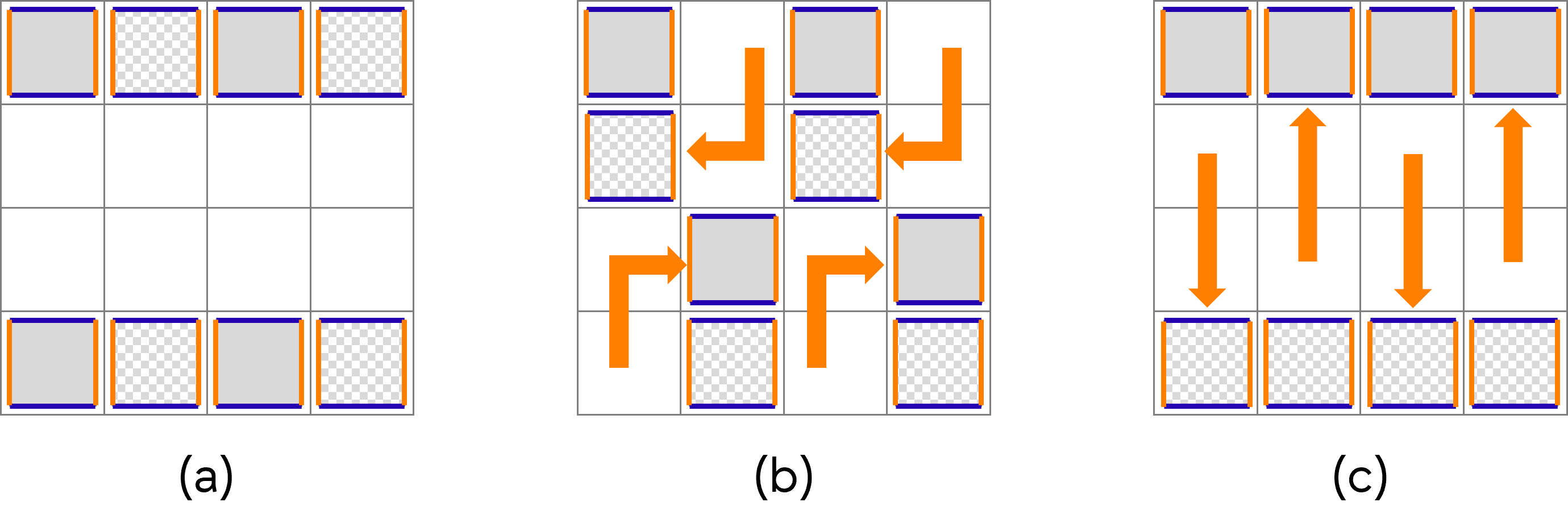}
    \caption{
    Lattice surgery operation of moving patches. 
    Filled (hatched) patches carry spin-up (spin-down) components. 
    (a) After completing the $ZZ$ rotation gates, patches are moved to split spin-up and spin-down components. 
    (b) Half of the logical patches are moved to the middle region. This operation takes 2 lattice surgery clocks. 
    (c) Then, Logical patches in the middle region are moved to the upper or lower side, depending on the spin components they carry. It takes 1 lattice surgery clock. 
    Therefore, the moving operation takes 3 logical clocks in total. 
    }
    \label{fig:movingpatch}
\end{figure}

Next, let us discuss how to perform the time evolution operator in lattice surgery term by term. 
We begin with the implementation of the $ZZ$ rotation terms, which are part of the second term in Eq.(\ref{eq:2Dhubbard_final}). 
The Pauli $ZZ$ operators in this term act on the spin-up and down components on the same lattice site. 
To perform these $ZZ$ rotations locally, we assign the spin components to logical patches as in Fig.~\ref{fig:init_Z_ZZ}. 
Figure~\ref{fig:ZZrot_ls} shows the actual lattice surgery operations needed for $ZZ$ interactions. 
Initially, each RUS process has two patches for injection at first (Fig.~\ref{fig:init_Z_ZZ}). 
As previously discussed, 
this initial injection region will be dynamically updated by the assignment algorithm. 
The execution time of the parallel $ZZ$ rotation is denoted as $T_{\rm RUS}(V, ZZ)$ clocks. 
We define the average parallel RUS execution clocks as $T_{\rm RUS}(M, P)$, where $M$ and $P$ are the number of RUS processes performed in parallel and the rotation basis Pauli operator, respectively, with $V = N^2$. 
After performing the $ZZ$ rotations, we move on to the more complicated operations: hopping interactions and fSWAP operations. 
But before that, we rearrange the logical patches to split the spin-up and down components into the upper and lower rows of the logical patches, respectively. 
This is necessary because the hopping terms are confined to each spin component and can be performed in parallel. 
This rearrangement is achieved by moving the logical patches, as shown in Fig.~\ref{fig:movingpatch}~\footnote{
Note that this rearrangement is just a modification of the position of the logical patch, not a modification of the JW ordering. 
Therefore, this operation does not need the fSWAP operations. 
}. 
This operation takes 3 clocks in total. 

\begin{figure}
    \centering
    \includegraphics[width=80mm, clip]{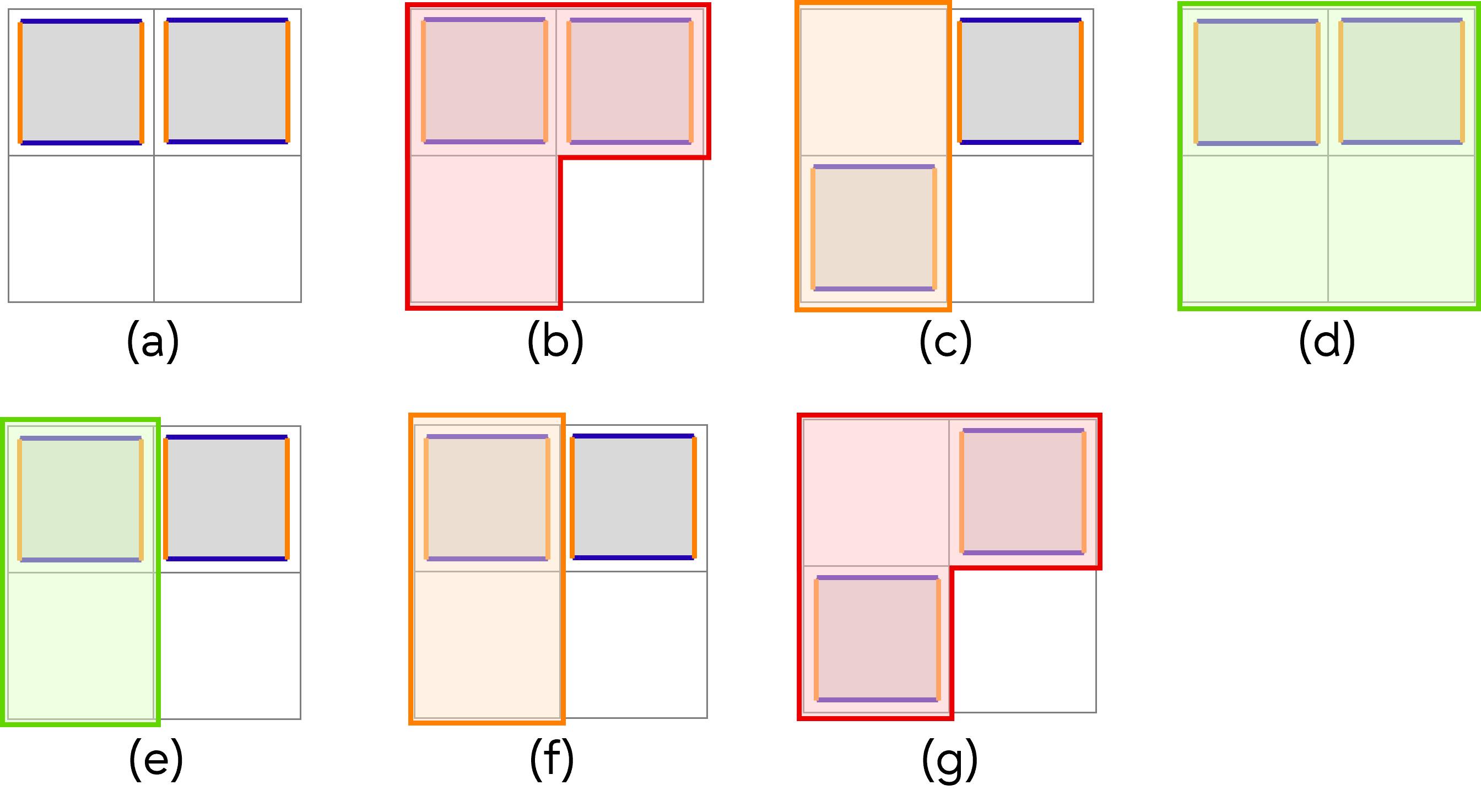}
    \caption{
    Lattice surgery operations of the $XX+YY$ rotation gate. Figures (a)-(g) are arranged by the order of lattice surgery operations. 
    The logical CNOT gate and the logical Hadamard gate are detailed in Appendix~\ref{appx:qecandls}.   
    (a) Initial configuration of the logical patches.
    (b) First logical CNOT operation by using the red quoted area. 
    (c) First logical Hadamard gate. 
    The logical patch that moved to the adjacent patch during the patch rotation does not need to move back to the original position. 
    Therefore, this logical Hadamard gate is performed in 2 lattice surgery clocks.
    (d) The RUS execution of the $ZZ$ rotation gate. The initial injection region is shown as the green area. 
    (e) The RUS execution of the single $Z$ rotation gate. 
    (f) Second logical Hadamard gate. Similarly to the first logical Hadamard gate, it takes only 2 lattice surgery clocks. 
    (g) Second logical CNOT gate. This operation takes 3 clocks due to the patch moving to the starting position. 
    The entire operation takes $9 + T_{\rm RUS}(M, ZZ) + T_{\rm RUS}(M', Z)$ lattice surgery clocks in total.  
    } 
    \label{fig:XXYYls}
\end{figure}

Next, let us move on to the lattice surgery implementation of the hopping interaction terms. 
We start by arranging the logical qubits in the first JW ordering, as shown in Figure~\ref{fig:fswap4x4} (ordering (a)). 
This set-up allows us to perform half of the hopping interactions via adjacent $XX + YY$ rotations. 
These interactions are further divided into two parts to avoid overlaps, as shown 
in Fig.~\ref{fig:parallelXXYY}. 
Each $XX+YY$ rotation is performed using the circuit in Fig.~\ref{fig:XXYYcircuit},  
with its lattice surgery operation schematically represented in Fig.~\ref{fig:XXYYls}. 
The number of $XX+YY$ rotations performed at the same time is $N(N-1)$ since each JW ordering contains $2N(N-1)$ local hopping terms, and they are divided into horizontal and vertical. 
Therefore, the execution time of the $XX+YY$ rotations in the first JW ordering is: 
$2 \times (9 + T_{\rm RUS}(N(N-1), ZZ) + T_{\rm RUS}(N(N-1), Z))$. 

\begin{figure}
    \centering
    \includegraphics[width=80mm, clip]{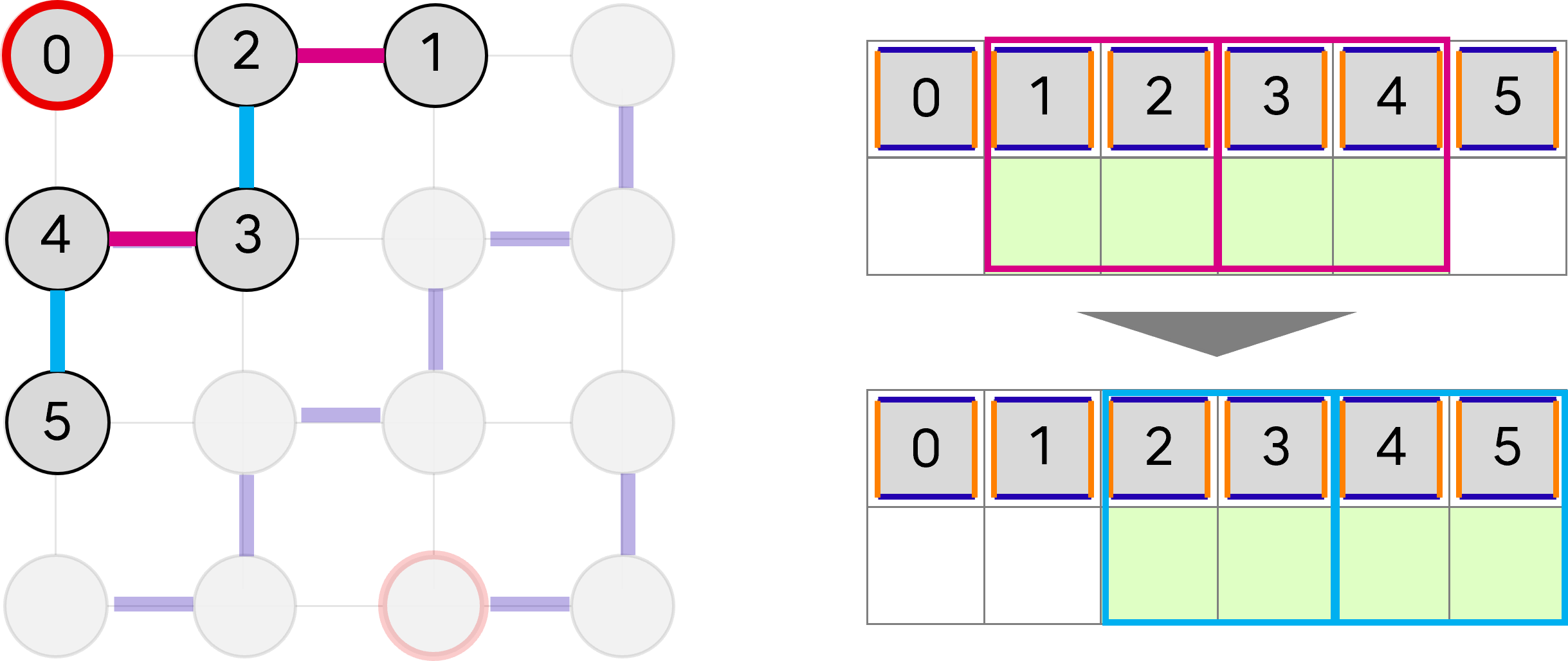}
    \caption{
    Example of the parallel $XX+YY$ rotation. 
    This figure shows a subset of up-spin hopping interaction terms of the $4 \times 4$ Hubbard model (deep color part of the left figure), but other parts can be performed in the same way. 
    The rotation gates are performed in two steps due to shared qubits: horizontal interactions (magenta edges) and vertical interactions (cyan edges). 
    The minimum RUS process is highlighted in the colored quoted area in the right figure. 
    Free patches at the beginning of the execution of the $XX + YY$ rotation (e.g. left-most white patches in the right figure) are assigned to the neighboring RUS process to boost the injection protocol. 
    }
    \label{fig:parallelXXYY}
\end{figure}
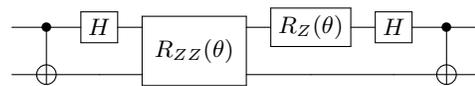
\begin{figure}
    \centering
    \mbox{
        \Qcircuit @C=1em @R=.7em {
        \lstick{} & \ctrl{1} & \gate{H} & \multigate{1}{R_{ZZ}(\theta)} & \gate{R_{Z}(\theta)} & \gate{H} & \ctrl{1} & \qw \\
        \lstick{} & \targ    & \qw      & \ghost{R_{ZZ}(\theta)}        & \qw                  & \qw      & \targ    & \qw 
        }
    }
    \caption{
    Circuit to implement $XX+YY$ rotation gate with a rotation angle $\theta$. 
    This circuit can be derived by the simultaneous diagonalization of $XX$ and $YY$ basis~\cite{reiher2017elucidating} or an equivalent transformation from the standard $XX+YY$ rotation circuit. 
    }
    \label{fig:XXYYcircuit}
\end{figure}

\begin{figure}
    \centering
    \includegraphics[width=80mm, clip]{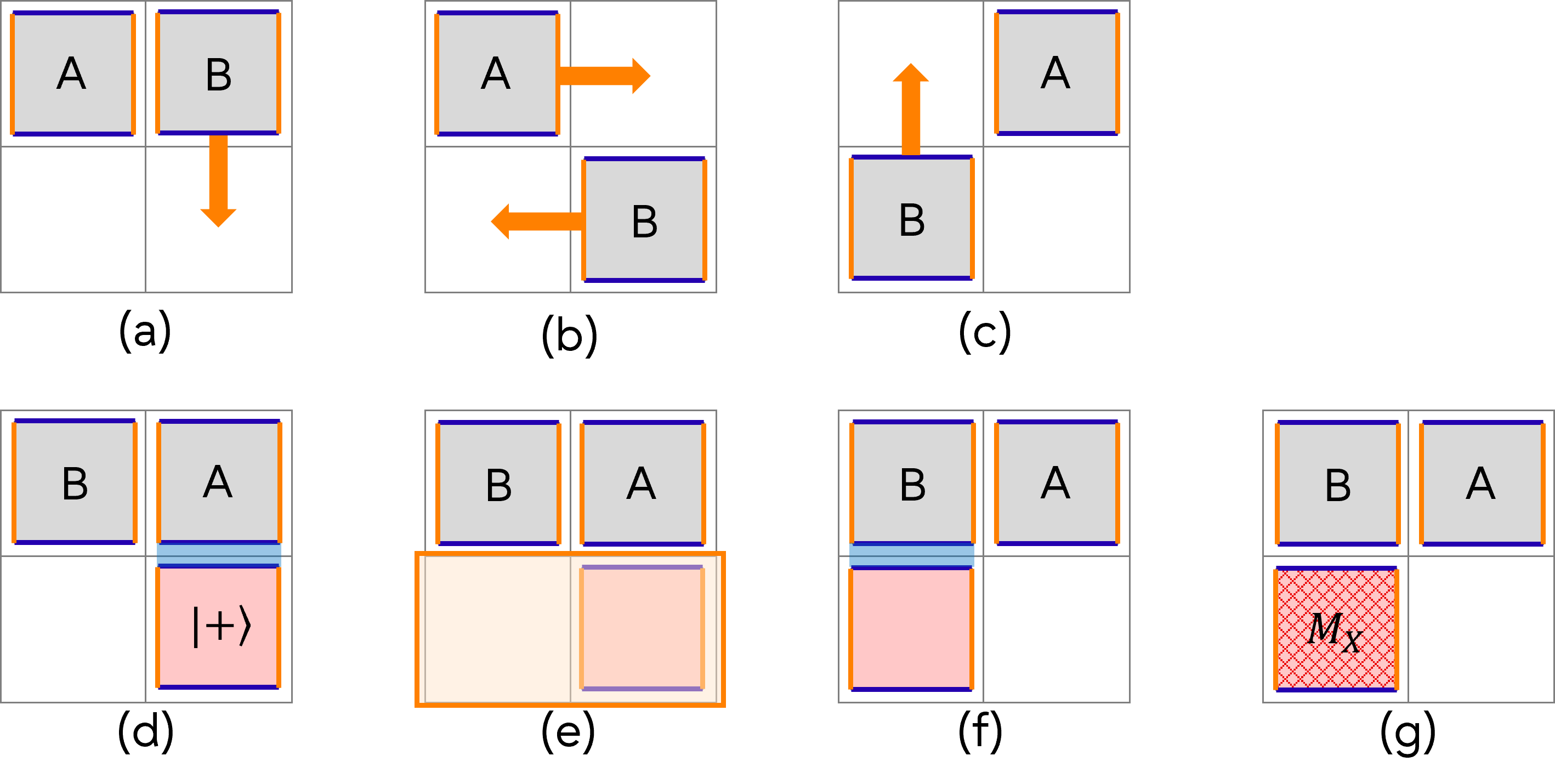}
    \caption{
    Lattice surgery implementation of the fSWAP gate. We perform the fSWAP gate by the sequence of the SWAP gate and the CZ gate. 
    (a) - (c) SWAP operation between patches A and B. 
    (d) - (g) Logical CZ operation between patches A and B. 
    (d) First, an ancilla patch is initialized to $\ket +$ state, then the $ZZ$ measurement is performed between patch A and the ancilla patch. 
    (e) Then, the logical Hadamard gate acts on the ancilla patch. When the patch rotates, the ancilla patch moves to the adjacent position. 
    (f) The ancilla patch and patch B are measured by the $ZZ$ operator. 
    (g) Finally, the ancilla patch is measured by the $X$ operator destructively. 
    In total, the fSWAP gate takes 7 lattice surgery clocks. 
    }
    \label{fig:fswapls}
\end{figure}

After completing the first half of the hopping terms, 
we switch to the other JW ordering using fSWAP operations. 
The logical fSWAP operation is realized through lattice surgery operations in Fig.~\ref{fig:fswapls}. 
We perform this logical fSWAP operation layer $N-1$ times for the $N \times N$ Hubbard model. 
The logical fSWAP operation takes 7 clocks, and therefore, the fSWAP layer between two JW orderings takes $7(N-1)$ clocks in total. 
Then, the remaining half of the hopping terms is performed similarly to the first JW ordering, taking $2 \times (9 + T_{\rm RUS}(N(N-1), ZZ) + T_{\rm RUS}(N(N-1), Z))$ clocks. 

\begin{figure}
    \centering
    \includegraphics[width=80mm, clip]{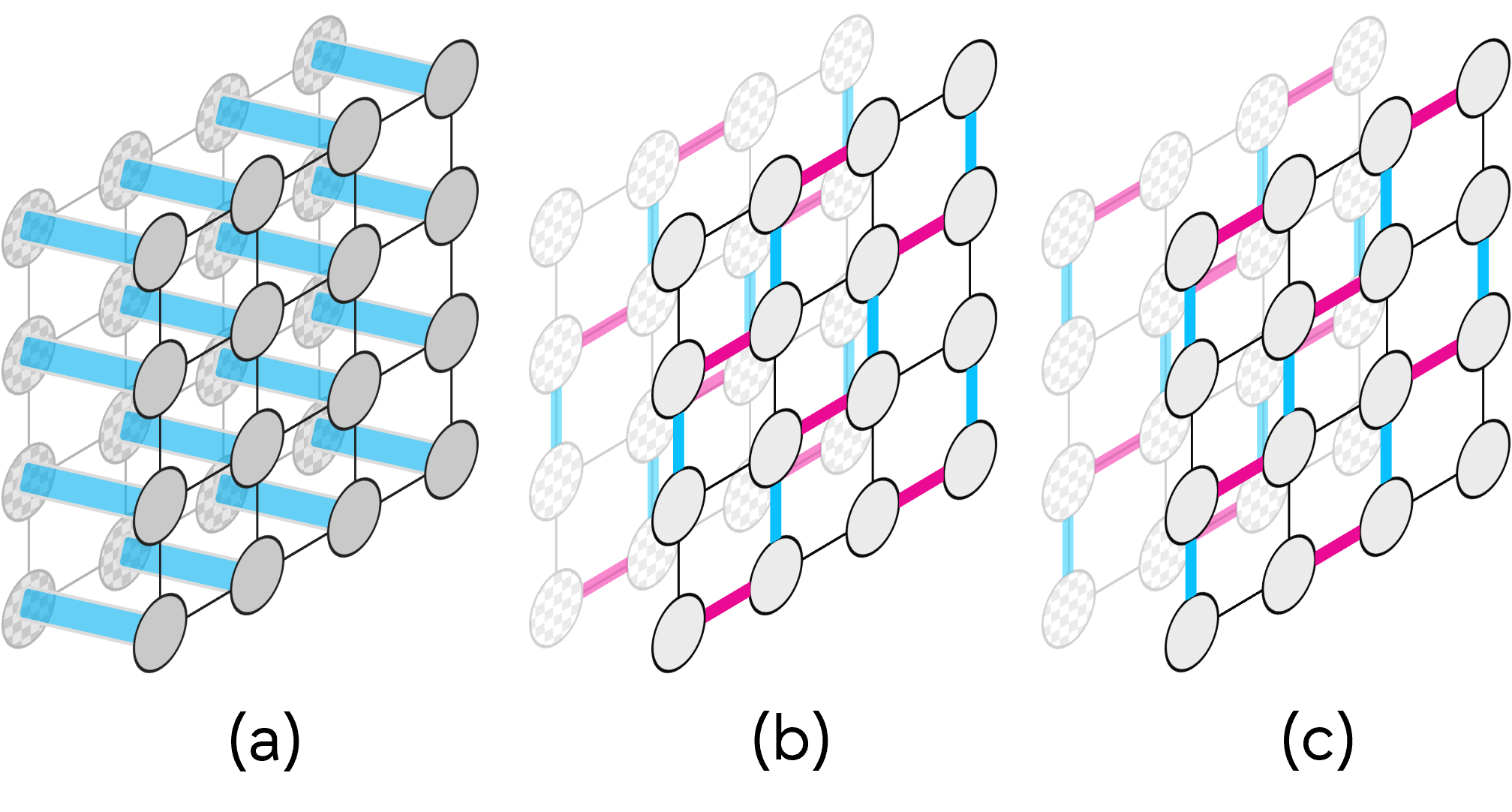}
    \caption{
    Order of operations in the Trotter time evolution for the $4 \times 4$ Hubbard model. 
    Interaction terms are performed in the order from (a) to (c). 
    Spin-up components are shown as filled circles, and spin-down components as hatched circles. 
    This figure illustrates the first half of the second-order Trotter formula. 
    The inverse order operations follow after operation (c). 
    (a) Inter-spin $ZZ$ rotation terms, followed by the patch moving layer. 
    (b) Neighboring $XX+YY$ rotation terms in the first JW ordering. As discussed before, interactions are divided into two subsets (magenta and cyan edges). 
    After this operation, we perform fSWAP layers to move on to the second JW ordering. 
    (c) $XX+YY$ rotation terms in the second JW ordering.
    }
    \label{fig:Trotter_order}
\end{figure}

Finally, let us estimate the execution time of the second-order Trotter time evolution. 
The order of the operations to perform the second-order Trotter time evolution is summarized in Fig.~\ref{fig:Trotter_order}.
Gathering the execution clocks of all the components, we obtain
\begin{equation} \label{eq:2nd_order_trott_clock}
    \begin{array}{ll}
        & T_{\rm Trotter} = 7T_{\rm RUS}(V-N, Z) + 7T_{\rm RUS}(V-N, ZZ) \\
        & + 2T_{\rm RUS}(V, ZZ) + 14N + 55.
    \end{array}
\end{equation}
The first two terms are derived from the RUS of $XX+YY$ rotation gates. 
Since the two $XX+YY$ rotation layers in the middle of the second-order Trotter formula can be merged into a single $XX+YY$ rotation layer, their coefficients become 7 instead of 8. 
By applying $T_{\rm Trotter}$ with an appropriate estimation of $T_{\rm RUS}(M, P)$, we can estimate the runtime of quantum algorithms using the time evolution operator in the STAR architecture. 
In Sec.~\ref{sec:phaseest}, we demonstrate the application to the quantum phase estimation algorithm and discuss the required resource in detail. 
Before delving into the resource estimation, we discuss how to estimate $T_{\rm RUS}(M, P)$ in the remainder of this section. 

\subsection{Estimation of $T_{\rm RUS}(M, P)$}
As discussed before, to estimate $T_{\rm RUS}(M, P)$, we consider the QEC parameters such as the physical error rate, 
the configuration of parallel RUS executions, 
and how the updating of the injection region during the execution. 
Therefore, in general, $T_{\rm RUS}(M, P)$ should be estimated by numerical simulations. 
In this study, 
we perform a two-step numerical simulation to estimate $T_{\rm RUS}(M, P)$:
the ancilla state injection and the parallel RUS execution. 
The former simulates the success rate of the injection protocol 
under the circuit-level noise model with a physical error probability $p_{\rm phys}$. 
This simulation uses a stabilizer formalism, and we employ Stim~\cite{Gidney_2021} as the backend. 
Next, we simulate the parallel RUS execution and the adaptive assignment algorithm based on the given injection success rate and the RUS configuration. 

\begin{figure*}
    \centering
    \includegraphics[width=80mm, clip]{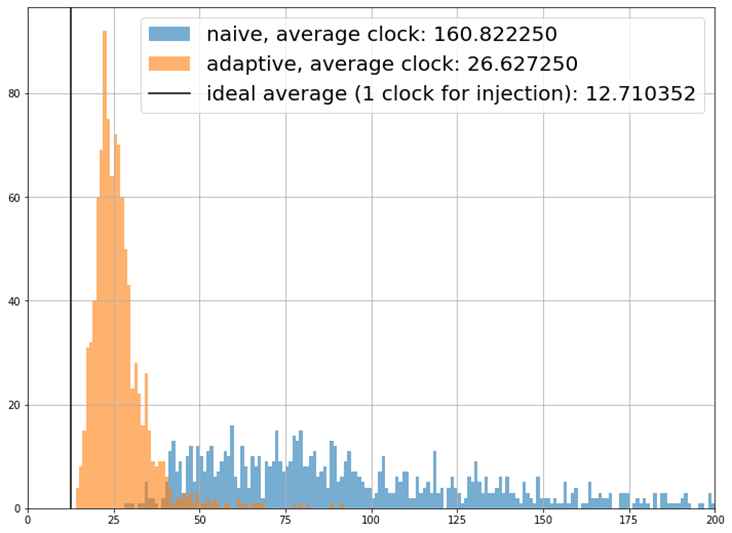} \hspace{10mm}
    \includegraphics[width=80mm, clip]{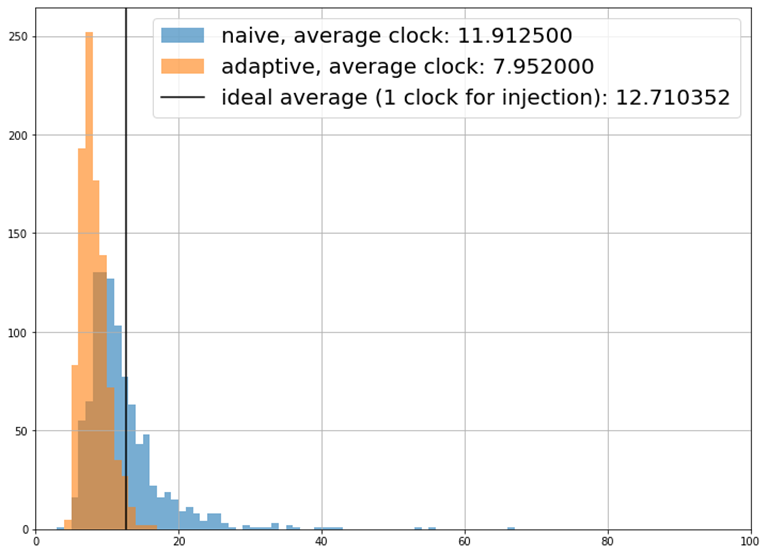}
    \caption{Histograms of the total RUS steps to finish 32 parallel RUS processes. The horizontal axis shows the runtime taken to complete all RUS processes in the unit of lattice surgery clock, while the vertical axis shows the event counts. 
    We also indicate the average runtime assuming that the injection succeeds with a single lattice surgery clock. 
    (left) Case of the physical error probability $p_{\rm phys} = 10^{-3}$.
    (right) Case of the physical error probability $p_{\rm phys} = 10^{-4}$.
    }
    \label{fig:parallelRUS_sim_hist}
\end{figure*}

As a benchmark, We simulate 32 parallel small-angle $Z$ rotations. 
The logical patch arrangement follows Fig.~\ref{fig:patch2nf2f}, and the initial injection region consists of a single adjacent patch for each RUS process, similar to Fig.~\ref{fig:updateinjectionregion} (a). 
We compare execution times with a fixed injection region (naive) versus the space-time parallel injection technique combined with the adaptive region updating (adaptive).
We run $10^3$ simulations for each case under two parameter sets: 
one corresponding to a low success rate ($p_{\rm phys} = 10^{-3}$ and $d = 16$) 
and the other one corresponding to a high success rate ($p_{\rm phys} = 10^{-4}$ and $d = 8$). 
The resultant histograms of these simulations are displayed in Fig.~\ref{fig:parallelRUS_sim_hist}. 

First, let us discuss the numerical results using the low success rate set-up shown in Fig.~\ref{fig:parallelRUS_sim_hist} (left). 
The long-lasting RUS processes, hindered by the low injection success rate, significantly degrade the performance of parallel execution.
This is evident from the blue histogram in Fig.~\ref{fig:parallelRUS_sim_hist} (left), which shows a long tail and an average runtime of about 161 clocks.
The majority of this time, almost 95\%, is consumed by the injection process, as the average RUS step count is less than 10 when $M = 32$, as shown in Fig.~\ref{fig:averageRUSparallel}, and the $ZZ$ measurement in the gate teleportation circuit takes only 1 clock.
Conversely, the parallel injection protocol and adaptive assignment algorithm 
significantly reduce this overhead, as shown by the orange histogram in Fig.~\ref{fig:parallelRUS_sim_hist} (left). 
The average runtime drops to about 27 clocks, achieving an 83\% reduction in execution time. 
This result confirms that parallel injection suppresses the injection overhead, while 
the adaptive assignment algorithm mitigates the overhead from the long-lasting RUS protocol. 
The very short tail in the orange histogram highlights the effectiveness of this cancellation, indicating that our adaptive parallel RUS execution is highly effective is scenarios with low injection rates.

For the high success rate set-up, the naive implementation faces fewer issues since injection trials rarely fail. 
However, our parallel injection and adaptive assignment algorithm improves the entire performance, as shown in the orange histogram in Fig.~\ref{fig:parallelRUS_sim_hist} (right). 
While the parallel injection does improve the success rate, 
the primary benefit 
comes from ``pre-injection'' for the next RUS trial during the $ZZ$ measurement of the current RUS trial, as illustrated in Fig.~\ref{fig:parallelinjection} (right). 
With injections almost always succeeding within 1 clock, 
this pre-injection technique hides the injection overhead of subsequent RUS trials. 
As a result, the average clock approaches the theoretical lower-bound value of $M=32$ (shown in Fig.~\ref{fig:averageRUSparallel}). 
In summary, our parallel injection and adaptive assignment algorithm is valuable across a wide range of scenarios, for both low and high injection success rates. 

If precise estimation of $T_{\rm RUS}(M, P)$ is not essential, numerical simulations may not be the most efficient approach.
Instead, one can use $\langle K \rangle_M$ values shown in Fig.~\ref{fig:averageRUSparallel} together with some assumptions about injection runtime.
The assumed injection clock should be an upper bound of the actual injection runtime. 
For example, if the physical error rate is as low as $p_{\rm phys} = 10^{-4}$, it is reasonable to assume that injection succeeds within 1 clock, a fact supported by the numerical verification in the resource estimation section (Sec.~\ref{sec:phaseest}).
Using this assumption, one can roughly estimate the average runtime of parallel $Z$ rotations as follows, 
\begin{equation} \label{eq:roughestofparallelRUSclock}
    \begin{array}{ll}
    T_{\rm RUS}(M, P) &\approx \langle K \rangle_M \cdot \left\{ (\text{1 clock for injection}) \right. \\
    & \left. + (\text{1 clock for $Z \otimes P$ measurement}) \right\}. 
    \end{array}
\end{equation}
By combining reasonable assumptions for the injection and Eqs. (\ref{eq:2nd_order_trott_clock}) and (\ref{eq:roughestofparallelRUSclock}), one can easily estimate the execution time of Trotter-based algorithms using the STAR architecture. 

\subsection{Comparison with the serial compilation strategy} 
Finally, let us compare our compilation result with a more standard approach in FTQC~\cite{Litinski2018latticesurgery}. 
In this standard compilation, the Clifford operations are absorbed into the rotation gates and Pauli measurements~\cite{Litinski2019gameofsurfacecodes}, allowing any quantum circuit to be executed using multi-qubit Pauli rotation gates and measurements.
Parallel execution of these multi-qubit Pauli rotations is generally challenging, so the rotation gates are executed sequentially (this compilation method is referred to as serial compilation).
We discuss the details of this compilation method in Appendix ~\ref{appx:serialcompile}. 
Figure~\ref{fig:vs_serial} shows the runtime of the second-order Trotter formula by using both parallel and serial compilations. 
The $T_{\rm RUS}(M, P)$ in the parallel compilation is estimated based on parameters discussed in Sec.~\ref{sec:phaseest}. 
As a result, we can achieve a significant reduction in runtime, ranging from 86\% to 97\% in this set-up. 
\begin{figure}
    \centering
    \includegraphics[width=80mm, clip]{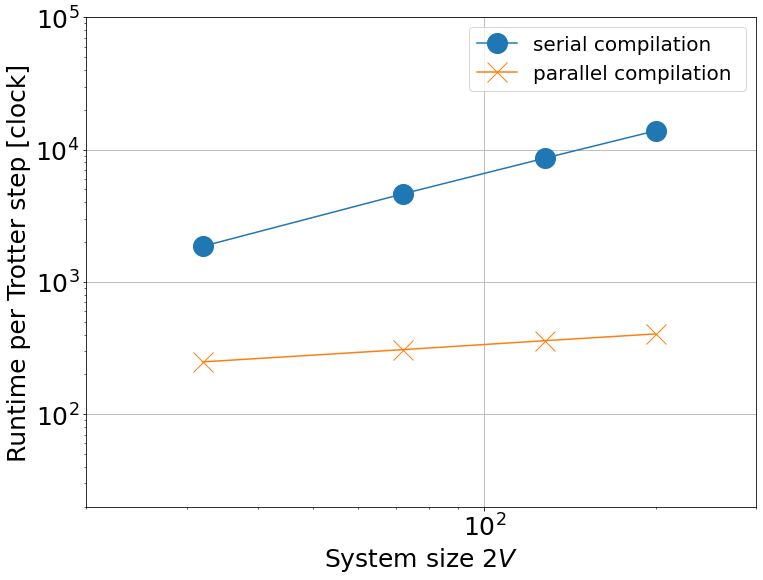}
    \caption{
    Comparison with the serial compilation result. 
    The horizontal axis indicates the system size $2V = 2M^2$ for the $M \times M$ Hubbard model. 
    Blue circles (orange crosses) indicate the runtime of a single Trotter step with the serial (parallel) compilation strategy. 
    }
    \label{fig:vs_serial}
\end{figure}

\section{Application of the compilation result} \label{sec:phaseest}
In this section, 
we demonstrate how to apply our compilation results to 
possible early-FTQC tasks, specifically the QPE algorithm. 
First, we introduce the QPE algorithm used in this study and 
estimate the resource requirements for the QPE of the 2D Hubbard model using our compilation results. 
While our estimation 
relies on empirical parameters from previous studies, 
we expect that the order of the estimated values does not change drastically. 
This demonstration will be useful for readers looking to estimate the resource requirements of certain algorithms within the STAR architecture.

\subsection{QPE algorithm for early-FTQC}
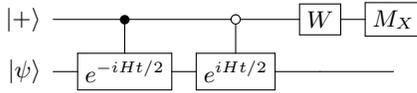
\begin{figure}
    \centering
    \mbox{
        \Qcircuit @C=1em @R=.7em {
        \lstick{\ket{+}}    & \ctrl{1}          & \ctrlo{1}           & \gate{W} & \gate{M_X}  \\
        \lstick{\ket{\psi}} & \gate{e^{-iHt/2}} & \gate{e^{iHt/2}}      & \qw      & \qw   \\
        }
    }    
    \caption{Hadamard test circuit. 
    We set $W = I$ ($W = S^{\dag}$) to calculate ${\rm Re} \bra{\psi} e^{-iHt} \ket{\psi}$ (${\rm Im} \bra{\psi} e^{-iHt} \ket{\psi}$). }
    \label{fig:hadamardtest}
\end{figure}

Focusing on early-FTQC devices, we employ a recently proposed QPE algorithm~\cite{PRXQuantum.4.020331} designed for such devices, rather than traditional textbook wisdom~\cite{textbook_QCQI}. 
This algorithm relies on the Hadamard test, as shown in Fig.~\ref{fig:hadamardtest}. 
The Hadamard test circuit requires only a single ancilla qubit, making it suitable for early-FTQC devices. 
To execute this circuit, we need to perform controlled time evolution operator. 
The anclla state in Fig.~\ref{fig:hadamardtest} is placed next to the routing region (Fig.~\ref{fig:ctrlpatch}).
In Appendix~\ref{appx:ctrlrot}, we discuss its implementation using our compilation results discussed above. 
By sampling and properly averaging the measurement results, we can extract the expectation value of the time evolution operator:  
\begin{equation}
    \bra{\psi} e^{-itH} \ket{\psi} = \sum_i p_i e^{-i E_i t},
\end{equation}
where $\ket{\psi} = \sum_i a_i \ket{E_i}$ is an input state, $\ket{E_i}$ is the $i$-th eigenstate of the Hamiltonian, and $p_i = |a_i|^2$. 
As observed, this expectation value is a sum of the exponential factors of different eigenstates, allowing us to extract energy eigenvalues from its time dependence using classical signal processing. 
In this study, we employ the quantum complex exponential least squares (QCELS) algorithm~\cite{PRXQuantum.4.020331} to achieve this. 
This algorithm is preferable since it requires relatively lower circuit depth compared to other algorithms~\cite{PRXQuantum.4.020331}. 
We cite the main algorithm of the QCELS algorithm and the statement of its computational complexity as Algorithm 1 and Theorem 2 in Appendix~\ref{appx:qcels}. The following resource estimation is based on these elements. 

\begin{figure}
    \centering
    \includegraphics[width=40mm, clip]{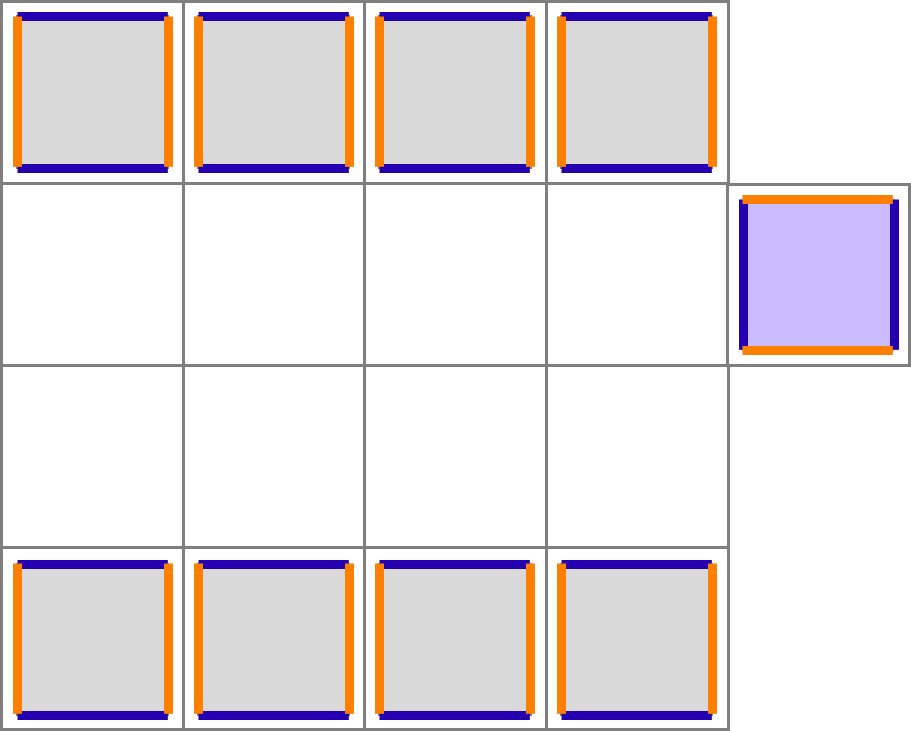}
    \caption{Patch arrangement in the QPE algorithm. 
    We show only a part of the entire patch arrangement. 
    The blue patch denotes the ancilla qubit for the Hadamard test circuit.}
    \label{fig:ctrlpatch}
\end{figure}

\subsection{Resource estimation}
Now we are ready to discuss the resource estimation details. 
First, we summarize some assumptions and parameter settings, such as the target precision. 
Based on these settings, we derive the required number of Trotter steps in the QCELS algorithm. 
By combining these results with our compilation results, 
we determine the code distance. 
After verifying this code distance through numerical simulations, 
we estimate the execution time and the number of physical qubits needed. 

We consider the ground state energy estimation with a certain precision $\epsilon_{\rm targ}$, $E_0 + \epsilon_{\rm targ}$, of the $M \times M$ Hubbard model ($M = 4,6,8,10$).  
The target precision of the QPE is set to $\epsilon_{\rm targ} = 0.01$. 
This choice is commonly used in previous studies~\cite{Kivlichan2020improvedfault,yoshioka2022hunting}.
Two algorithmic errors affect the final estimation: the Trotter error $\epsilon_{\rm Trotter}$ and the estimation error of the QCELS, $\epsilon_{\rm QCELS}$. 
In the worst case, their linear combination contributes to the error of the final result, so they should satisfy: 
\begin{equation} \label{eq:prec_criterion}
    \epsilon_{\rm QCELS} + \epsilon_{\rm Trotter} \leq \epsilon_{\rm targ}.
\end{equation}
As seen in Theorem 2, the QCELS algorithm assumes the target phase is in $[-\pi, \pi)$, so we normalize the Hamiltonian as follows,
\begin{equation}
    \tilde H \equiv \frac{\pi}{\lambda} H, \quad \lambda \equiv ||H||_{\rm 1},
\end{equation}
where $|| H ||_{\rm 1} = \sum_j |c_j|$ is the sum of absolute values of the coefficients in the Hamiltonian (see Eq.(\ref{eq:Hamiltonian_paulidecomp})), usually referred to as the 1-norm of the Hamiltonian. 
Therefore, the terms in Eq.(\ref{eq:prec_criterion}) are replaced as follows, 
\begin{equation} \label{eq:prec_criterion_norm}
    \tilde \epsilon_{\rm QCELS} + \tilde \epsilon_{\rm Trotter} \leq \tilde \epsilon_{\rm targ}, 
\end{equation}
where tilde denotes that it is normalized by the factor of $\pi / \lambda$. 
The $\lambda$ values can be estimated for each $M \times M$ Hubbard model and listed in Tab.~\ref{tab:lambda}. 
\begin{table} 
    \centering
    \begin{tabular}{c|cccc}
         & $4 \times 4$ & $6 \times 6$ & $8 \times 8$ & $10 \times 10$  \\ \hline \hline
        $\lambda$ & 64 & 156 & 288 & 460 \\
    \end{tabular}
    \caption{$\lambda$ of our Hamiltonian.}
    \label{tab:lambda}
\end{table}
The QCELS algorithm parameters are determined by the statement of Theorem 2 in Ref.~\cite{PRXQuantum.4.020331}: 
$J=\left\lceil\log_2(1/ \tilde \epsilon_{\rm QCELS})\right\rceil+1$, 
$ \tau_j=2^{j-J} \frac{\delta}{N \tilde \epsilon_{\rm QCELS}}, \forall 1\leq j\leq J$. 
The choice of $\delta$ is somewhat empirical. 
Theoretically, $\delta$ is chosen as $\delta = \Theta(\sqrt{1 - p_0})$, where $p_0$ is the overlap factor between the input state and the ground state. 
However, numerical simulations in Ref.~\cite{PRXQuantum.4.020331} indicate that $\delta$ can be smaller than the theoretical prospect, namely $\delta \approx 0.06$ with $p_0 = 0.8$ (note that $\sqrt{1-0.8} \approx 0.45$). 
We optimistically use this value in our estimation. 
We set the number of data pairs $N = 5$  and the number of samples to $N_s = 100$. 
This choice is consistent with the numerical simulations shown in Ref.~\cite{PRXQuantum.4.020331}.
This parameter setting aligns with the leading behavior (except for the polylog factor) of the statement of Theorem 2 as $N N_s \approx 1 / \delta^2$.

Next, let us discuss $\tilde \epsilon_{\rm Trotter}$. In this estimation, we employ the second-order Trotter decomposition. 
We split each $\tau_j / 2$ ($j = 1,2,\dots,J$) into $N_j$ Trotter steps (note that the factor of 2 comes from the Hadamard circuit in Fig.~\ref{fig:hadamardtest}), $\Delta \tau_j = \tau_j / 2 N_j$. 
From Eq.(\ref{eq:2nd_trott_err}), $\Delta \tau_j$ must satisfies,
\begin{equation}
    \tilde W \Delta \tau_j^2 \leq \tilde \epsilon_{\rm Trotter}, 
\end{equation}
and therefore $N_j$ should satisfies,
\begin{equation} \label{eq:trotter_step}
    N_j \geq \frac{\tau_j}{2} \sqrt{\frac{\tilde W}{\tilde \epsilon_{\rm Trotter}} }, 
\end{equation}
where $\tilde W$ is a rescaled Trotter error norm defined as $\tilde W = (\pi/\lambda)^3 \cdot W$. 
The $\tilde W$ value should be estimated to determine $N_j$. 
We use the $W$ value provided in Ref.~\cite{Kivlichan2020improvedfault} in our estimation and re-scale it (although details of the set-up are different from Ref.~\cite{Kivlichan2020improvedfault}, we expect that the value of $W$ is similar). 

By combining Eqs. (\ref{eq:t_max}), (\ref{eq:t_total}), and (\ref{eq:trotter_step}), the total number of Trotter steps is derived as follows: 
\begin{equation} \label{eq:total_trotter_step}
    N_{\mathrm{total}}=\sum^J_{j=1} \frac{(N-1)N_s 2^{j - J(\tilde \epsilon_{\rm QCELS}) - 1} \delta}{\tilde \epsilon_{\rm QCELS}} \sqrt{\frac{\tilde W}{\tilde \epsilon_{\rm Trotter}}}. 
\end{equation}
We can minimize $N_{\rm total}$ by optimizing $\tilde \epsilon_{\rm QCELS}$ and $\tilde \epsilon_{\rm Trotter}$ under the condition of Eq.(\ref{eq:prec_criterion_norm}). 
Using the optimization results, 
we can calculate the maximal number of Trotter steps needed for a single Hadamard test circuit, $N_{\rm max} = N \cdot N_J = \frac{\delta}{2 \tilde \epsilon_{\rm QCELS}} \sqrt{\frac{\tilde W}{\tilde \epsilon_{\rm Trotter}}}$. 
These values are listed in Tab.~\ref{tab:NtotandNmax}.
\begin{table} 
    \centering
    \begin{tabular}{c|cccc}
         & $4 \times 4$ & $6 \times 6$ & $8 \times 8$ & $10 \times 10$  \\ \hline \hline
        $N_{\rm total}$ [Trotter step] & 2717609 & 4040743 & 5399835 & 6830416 \\
        $N_{\rm max}$   [Trotter step] & 3397 & 5051 & 6750 & 8538
    \end{tabular}
    \caption{Resultant $N_{\rm total}$ and $N_{\rm max}$ by the minimization of Eq.(\ref{eq:total_trotter_step}).}
    \label{tab:NtotandNmax}
\end{table}

From this, we determine the required code distance in two steps: 
(i) we estimate the total number of logical operations, $N_{\rm op}$, from $N_{\rm max}$ and $T_{\text{Trotter}}$ (Eq.(\ref{eq:2nd_order_trott_clock})), and then
(ii) determine the code distance that satisfies $p_{\rm logical}(d) \cdot N_{\rm op}(d) < \epsilon_{\text{logerr}}$, where $p_{\rm logical}$ is the logical error rate per single logical operation, and $\epsilon_{\text{logerr}}$ is the required logical error rate in a single circuit run. 
In the following estimation, we employ $p_{\rm logical} = 0.1 d (100 p_{\rm phys})^{\frac{d+1}{2}}$ with the physical error rate $p_{\rm phys} = 10^{-4}$, while $\epsilon_{\rm logerr}$ is set to 0.01~\cite{Litinski2019gameofsurfacecodes}. 
The $N_{\rm op}$ depends on the code distance $d$ via the success rate of the ancilla state injection in the RUS process. 
As discussed in Sec.~\ref{sec:compile}, the success rate and average execution time of RUS should be estimated numerically for precise results. 
However, because running numerous simulations for various code distances is inefficient, we first assume that the injection finishes within 1 clock, and roughly determine the code distance. 
The assumption is expected to closely match the actual scenario, given the physical error rate 
$p_{\rm phys}=10^{-4}$. 
After determining the code distance, we numerically simulate the parallel RUS execution. 
If the numerical runtime is smaller than the rough estimation, the chosen code distance is sufficient. 
The detemined code distance and estimated runtime per Trotter step are 
summarized in Tabs.~\ref{tab:injectionparams} and \ref{tab:clockperTrotterstep}, respectively. 
\begin{table*}
    \centering
    \begin{tabular}{c|cccc}
                         & $4 \times 4$ & $6 \times 6$ & $8 \times 8$ & $10 \times 10$  \\ \hline \hline
         $d$             & 9            & 11           & 11           & 11              \\
         $|Q_i|$  & $[3,3,3]$ ($k=3$) & $[2,2,2,2,3]$ ($k=5$) & $[2,2,2,2,3]$ ($k=5$) & $[2,2,2,2,3]$ ($k=5$) 
    \end{tabular}
    \caption{Verified code distance and parameter of the injection protocol.}
    \label{tab:injectionparams}
\end{table*}
\begin{table}
    \centering
    \begin{tabular}{c|cccc}
        & $4 \times 4$ & $6 \times 6$ & $8 \times 8$ & $10 \times 10$  \\ \hline \hline
        $T_{\rm Trotter}$ [clock] & 248.355 & 307.51 & 359.51 & 404.25  \\
    \end{tabular}
    \caption{Average clock per Trotter step derived from numerical simulations.}
    \label{tab:clockperTrotterstep}
\end{table}

\begin{figure}
    \centering
    \includegraphics[width=80mm, clip]{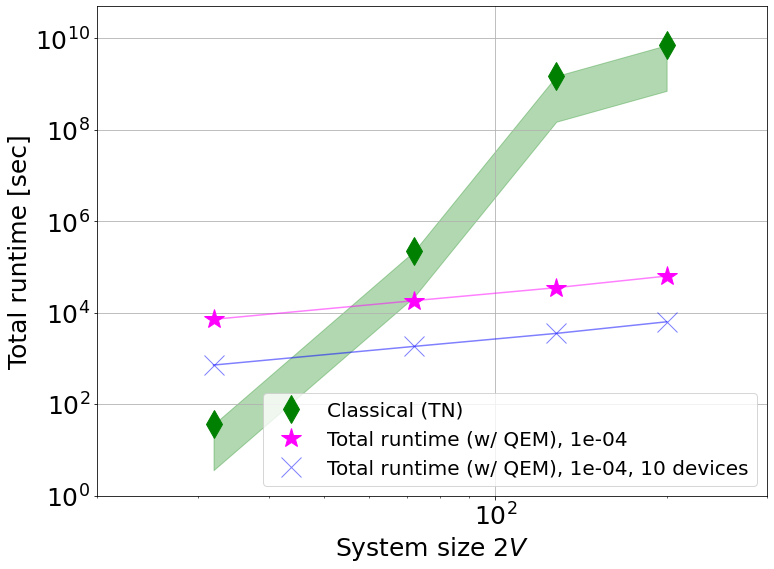}
    \caption{
    Comparison of total runtime for the QPE algorithm with that for classical tensor network (TN) calculations. 
    The horizontal axis indicates the system size $2V = 2M^2$ for the $M \times M$ Hubbard model. 
    Green plot shows the runtime of the classical algorithm estimated in Ref.~\cite{yoshioka2022hunting}.
    Blue plot displays the possible total runtime using 10 independent quantum devices in our calculation.}
    \label{fig:vsclassical}
\end{figure}

Once the code distance and average RUS execution time are determined, we can estimate the total execution time in seconds, assuming each code cycle (= syndrome measurement) takes 1 $\mu s$~\cite{Litinski2019gameofsurfacecodes}. 
To make our estimation more precise, we also account for the PEC overhead to compensate for the logical errors from the injection. 
By combining the second-order Trotter formula of Eq. (\ref{eq:2nd_trott_formula}) and the PEC overhead formulae of Eqs.(\ref{eq:ver2_wcler}) and (\ref{eq:pec_overhead}), we calculate the sampling overhead of the Hadamard test with time evolution $\tau$ as 
\begin{equation} \label{eq:qecoverhead_htest}
    C_{\tau} \approx e^{4 \alpha_{\rm RUS} \pi \tau p_{\rm phys}}. 
\end{equation}
We estimate the overhead of Eq.(\ref{eq:qecoverhead_htest}) for all Hadamard test circuits with $\{ \tau = \tau_j n /2 \mid j=1,\dots J,\quad  n = 0,1,\dots,N-1 \}$ and include them in the total runtime estimation. 
The number of physical qubits is estimated as $2d^2$ for each surface code patch. 
Since our patch arrangement needs $4M^2 + 1$ logical patches for $M \times M$ Hubbard model simulation ($+1$ comes from the ancilla qubit of the Hadamard test circuit), 
the total number of physical qubits is given as $(4M^2 + 1) \times 2 d^2$ for the $M \times M$ Hubbard model. 
These results are summarized in Tab.~\ref{tab:finalestimationresult}. 
In Fig.~\ref{fig:vsclassical}, we compare our total runtime estimation with that of a classical computer using the tensor network algorithm~\cite{yoshioka2022hunting}. 
Although the boundary conditions in the classical set-up in Ref.~\cite{yoshioka2022hunting} differ from ours, we expect the runtime order to be similar.
Our rough estimation indicates the possibility for quantum speedup in the $8 \times 8$ Hubbard model simulation, even with $6.5 \times 10^4$ physical qubits. 
We also see the potential in the 
$6 \times 6$ Hubbard model, although this is less certain owing to the 
overhead of state preparation in the QPE, which we did not consider in our estimation. 
For the $4 \times 4$ Hubbard model the runtime is slower than the classical algorithm, but it completes in about two hours and requires only about $10^4$ physical qubits.
This set-up is valuable for the early-FTQC era 
to directly compare quantum and classical algorithm results, verifying the quantum computer's accuracy. 
Given the recent progress in quantum computing hardware, 
such verification is expected to be possible in the near future. 
This result indicates that our architecture can 
serve as a good testbed for achieving quantum speedup in meaningful tasks during the early-FTQC era.

Finally, we discuss the possibility for runtime reduction. 
The total runtime presented assumes that a single quantum computer with over $10^4$ physical qubits runs all the necessary Hadamard test circuits in series. 
However, since these circuit runs are independent, we can easily reduce the total runtime by running them in parallel using multiple quantum computers. 
For example, if we employ 10 independent quantum computers, the possible total runtime is represented by the blue points in Figure~\ref{fig:vsclassical}. 
The maximum runtime shown in Tab.~\ref{tab:finalestimationresult} represents the theoretical lower bound achievable through a parallel sampling. 
Generally, once a quantum device with order of $10^4$ physical qubits is available, producing 10 such devices seems much easier than developing a single device with $10^5$ physical qubits that operates coherently. 
Typical FTQC algorithms require more physical qubits to operate coherently to reduce runtime by increasing distillation blocks. 
Therefore, achieving quantum speedup based on our proposed parallel sampling approach is promising, even from the perspective of device development. 
\begin{table}[tbp]
    \centering
    \begin{tabular}{c|cccc}
        & $4 \times 4$ & $6 \times 6$ & $8 \times 8$ & $10 \times 10$  \\ \hline \hline
        total runtime [sec] & 7158.25 & 18313.99 & 35246.92 & 63219.87 \\
        max runtime [sec] & 7.59 & 17.09 & 26.69 & 37.97 \\
        $N_{\rm qubit}$     & 10530 & 35090 & 62194 & 97042   
    \end{tabular}
    \caption{Estimated runtime of the QPE algorithm, and the required number of physical qubits with $p_{\rm phys} = 10^{-4}$.
    Total runtime includes all sampling runtimes of the Hadamard test circuit accounting for the QEM overhead. 
    Max runtime refers to the maximum runtime of the single run of the Hadamard test circuit, as estimated by $N_{\rm max}$ in Tab.~\ref{tab:NtotandNmax}. }
    \label{tab:finalestimationresult}
\end{table}

\section{Summary} \label{sec:summary}
In this paper, we focus on simulating the 2D Hubbard model and 
discuss how to effciently perform its time evolution operator on the STAR architecture. 
We use the fSWAP gate to localize interaction terms, 
enabling parallel execution of local rotation gates. 
To reduce the overhead from ancilla state injection, 
we combine two strategies: space-time parallel injection and adaptive updating of the injection region. 
These approaches result in over an 80\% reduction in the execution time against the naive serial rotation gate scheme, especially when the injection success rate is low.
Using the compilation results of the time evolution operator, we estimate the resource requirements for the QPE algorithm designed for early-FTQC devices. 
Our rough estimation indicates that we can achieve over $10^3$ times quantum speedup for simulating the $8\times 8$ Hubbard model using $6.5 \times 10^4$ physical qubits with $p_{\rm phys} = 10^{-4}$. 
This 
is encouraging, as it suggests that realizing quantum speedup for practical tasks may happen sooner than previously expected, thanks to the STAR architecture. 

Finally, we comment on some topics we do not address in this paper. 
(i) Other possible applications of the compilation result: 
The compilation result we obtained is a simple time evolution operator, which can be applied to several tasks beyond the QPE. 
For example, the controlled time evolution operator can be used for the imaginary time evolution~\cite{PhysRevResearch.4.033121}, which is essential for probing the systems's finite-temperature structure and increasing the ground state overlap. 
In addition, we can estimate the runtime for variational ansatzes with similar structures, 
such as the variational Hamiltonian ansatz (VHA) of the 2D Hubbard model. 
Combining the VHA with 
the local variational quantum compiling (LVQC)~\cite{PRXQuantum.3.040302,PhysRevResearch.5.033070} can further reduce the circuit depth. 
Additionally, the runtime for the quantum-selected configuration interaction (QSCI) algorithm~\cite{kanno2023quantumselected} can be estimated by treating the variational ansatz as the adiabatic state preparation circuit.
(ii) Compilation of other systems: 
While this paper focuses on the 
2D Hubbard model Hamiltonian, 
our compilation strategy can be applied to other systems. 
An important class of Hamiltonian is provided as~\cite{PhysRevX.8.011044}, 
\begin{equation} \label{eq:general_Hamiltonian}
    H = \sum_{p,q} T_{pq} c^{\dag}_p c_q + \sum_{p} U_p n_p + \sum_{p \neq q} V_{pq} n_p n_q.
\end{equation}
This Hamiltonian contains a range of physical systems, 
such as molecules with a discretized space cell and crystalline solids using the plane wave basis~\cite{PhysRevX.8.011044}. 
Using the fSWAP network, 
we can achieve the Trotter decomposition of the time evolution of Eq.(\ref{eq:general_Hamiltonian}) 
with a linear-depth circuit~\cite{PhysRevLett.120.110501}. 
Our parallel RUS execution technique further allows for the parallel execution of rotation gates $e^{-i T_{pq} (c^{\dag}_p c_q + c^{\dag}_q c_p) \Delta \tau }$ and $e^{-iV_{pq} n_p n_q \Delta \tau}$ that appear for each fSWAP layer. 
(iii) Combining other QEC codes: 
Our injection protocol can be applied to 
any CSS code beyond the surface code, 
offering room for improvement in the choice of the QEC code. 
Successfully integrating recently proposed QLDPC code families into our framework could reduce resource requirements further. 
While this would require changes in the compilation, the basic strategy remains the same: 
parallel RUS execution with efficient suppression of the injection overhead. 

We hope that our results stimulate discussions about applications in the early-FTQC era and accelerate the realization of practical quantum advantages.

\vspace{5mm}
NOTE ADDED: After completing the manuscript, we received a private communication from the authors of Ref~\cite{yoshioka2022hunting}. They claimed that, assuming $p = 10^{-4}$ in their estimation, the one-level distillation protocol becomes sufficient and the total number of physical qubits becomes less than $10^5$, which might lead to better or comparable performance than our framework. This is related to the observation that qubitization approach is significantly outperforms the Trotter approach for the phase estimation of 2D Hubbard model, as shown in Ref.~\cite{yoshioka2022hunting}. However, if considering tasks where the Trotter approach performs comparably or better than qubitization approach, such as Hamiltonian simulations~\cite{Childs_2018}, our framework will offer substantially greater reductions in spacetime resources compared to full-fledged FTQC.
Further exploration should be needed to understand when and how our approach provides advantage against Clifford + $T$ approach.

\begin{acknowledgments}
K.F. is supported by MEXT Quantum Leap Flagship Program (MEXT Q-LEAP) Grant No. JPMXS0120319794, 
JST COI-NEXT Grant No. JPMJPF2014, and JST Moonshot R\&D Grant No. JPMJMS2061.
\end{acknowledgments}


\appendix

\section{Surface code and Lattice surgery} \label{appx:qecandls}
In this section, we summarize the basics of the rotated surface code and lattice surgery used in this study.

\subsection{Definition of the rotated surface code}
\begin{figure}[tbp]
  \centering
  \includegraphics[width=40mm, clip]{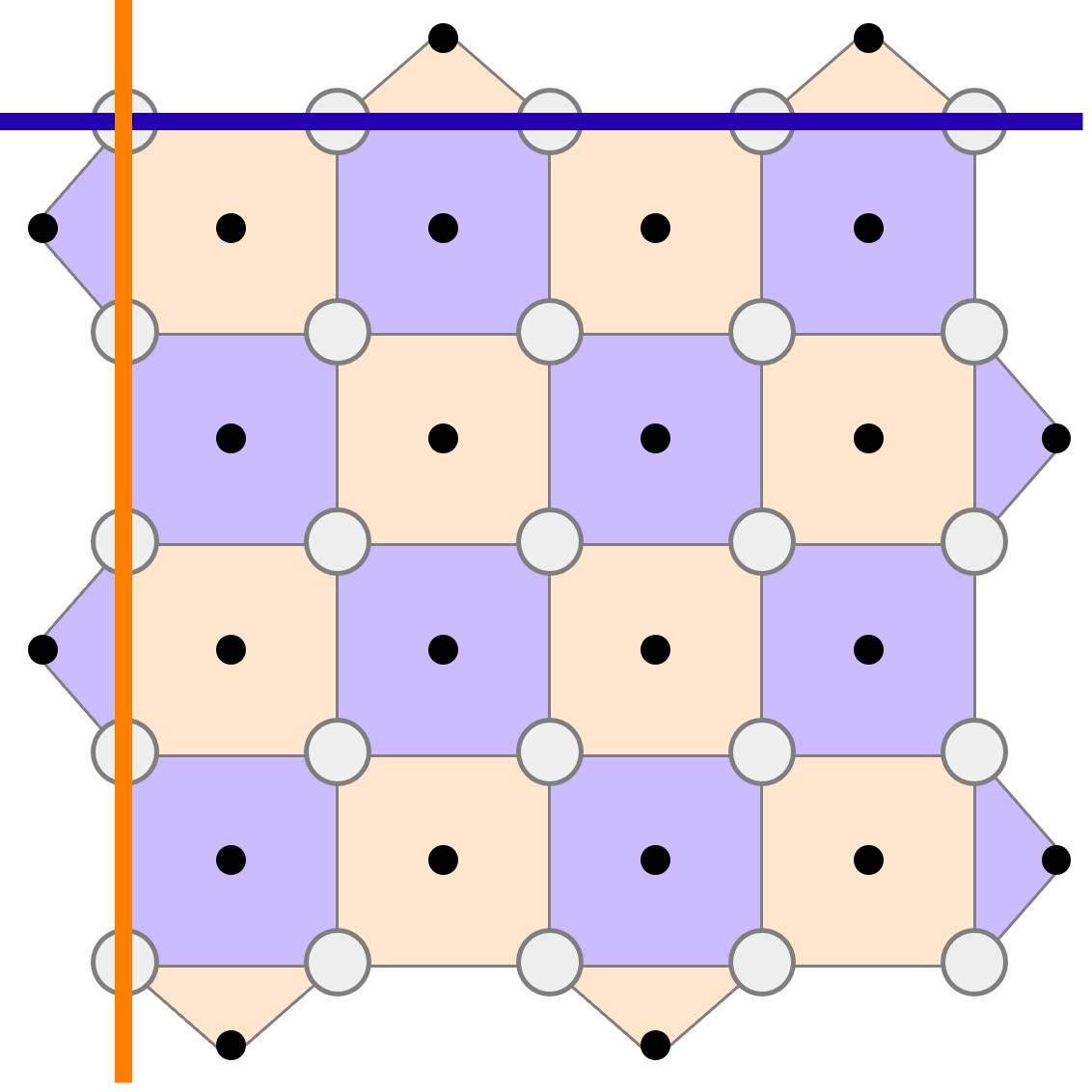}
  \caption{Rotated surface code with code distance $d = 5$. 
  White and black circles represent physical qubits used for encoding and measurement, respectively. 
  $X$ ($Z$) stabilizer operators are shown as orange (blue) faces. 
  Representative logical $X$ ($Z$) operators are given in orange (blue) solid lines on the boundary.}
  \label{fig:surface_code_def}
\end{figure}
\begin{figure}[tbp]
  \centering
  \begin{tabular}{cc}
    \includegraphics[width=22mm, clip]{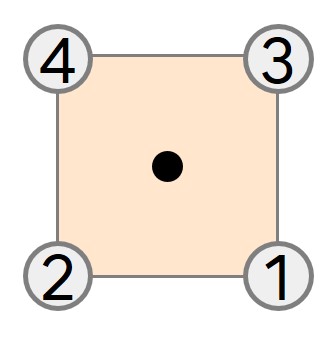} &
    \includegraphics[width=40mm, clip]{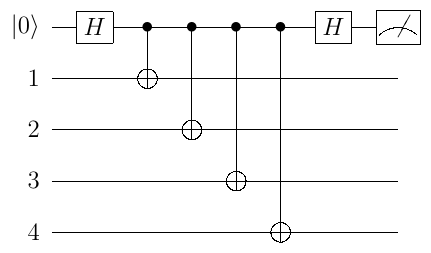} 
    \\ 
    \includegraphics[width=22mm, clip]{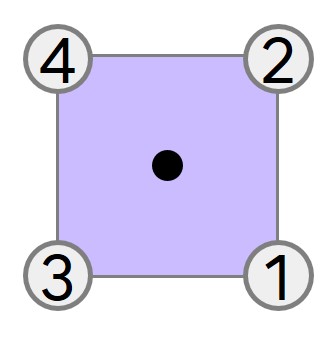} & 
    \includegraphics[width=40mm, clip]{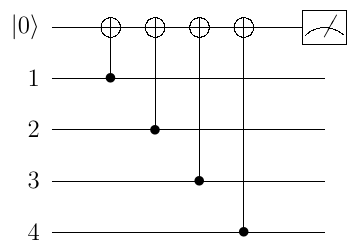}   
  \end{tabular}
  \caption{Syndrome measurement circuits. 
  (Upper) $X$ syndrome measurement circuit, with the order of CNOT operations represented by circled numbers on the left-hand side. 
  (Lower) $Z$ syndrome measurement circuit.}
  \label{fig:meas_circ}
\end{figure}
Fig.~\ref{fig:surface_code_def} illustrates the rotated planar surface code. 
This code is defined on physical qubits arranged in a $d \times d$ lattice (represented by white circles in Fig.~\ref{fig:surface_code_def}). The stabilizer operators are defined on the faces of the lattice (shown as orange and blue faces in Fig.~\ref{fig:surface_code_def}). 
The measurement circuit of the stabilizer operator and the order of the CNOT operation are provided in Fig.~\ref{fig:meas_circ}, with ancilla qubits used in the measurement circuit depicted as black dots in Fig.~\ref{fig:surface_code_def}). 
Logical operators are defined on the lattice boundary, indicated by orange and blue solid lines. 
The minimal length of the logical operator matches the side length of the lattice, making the code distance $d$. 

\subsection{Surface code lattice surgery}
The surface code inherently supports logical transversal CNOT gates owing to its CSS code strcture. 
However, in practical scenarios where entangling gates are limited to certain pairs of physical qubits, such as the 2D nearest-neighbor connectivity in superconducting devices, transversal CNOT gates are not feasible. 
The clever way to perform logical FT gates in such a restricted scenario is known as lattice surgery~\cite{Horsman_2012,Litinski2019gameofsurfacecodes}. 
Lattice surgery requires only the measurement of local stabilizer operators, which can easily be achieved even with restricted connectivity. 

The fundamental lattice surgery operations include patch merging, patch splitting, and patch deformation. 
By combining these fundamental operations with additional specific techniques for certain logical gates, 
we can perform all logical Clifford gates fault-tolerantly. 
Below, we introduce how to perform $H, CNOT, S$ gate in this study. 

\subsubsection{Hadamard gate}
\begin{figure}
    \centering
    \includegraphics[width=80mm, clip]{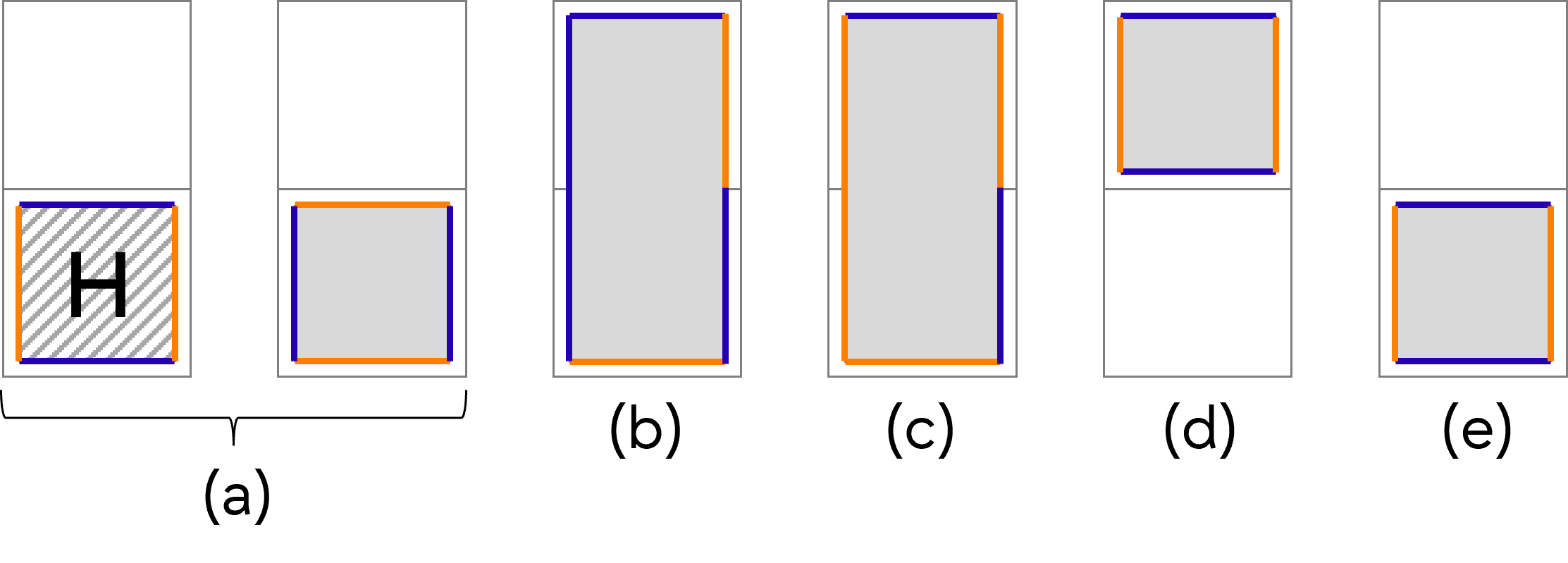}
    \caption{Logical Hadamard gate using lattice surgery.}
    \label{fig:hadamard_ls}
\end{figure}

For the rotated surface code, the logical Hadamard gate can be performed transversally. 
However, applying the transversal Hadamard gate rotates the patch surface by 90 degrees (Fig.~\ref{fig:hadamard_ls} (a)). 
To change the patch direction to its original one, we first deform the patch into a rectangular shape with a modified boundary (Fig.~\ref{fig:hadamard_ls} (b)). 
Then, we adjust the boundary and shrink the rectangular patch into a square shape (Fig.~\ref{fig:hadamard_ls} (c)-(e)). 
Although this procedure brings the patch almost back to its original position, 
it slightly differs by two lines of physical qubits (not shown in Fig.~\ref{fig:hadamard_ls}). 
To remedy this small shift, we use the ``patch sliding'' technique~\cite{McEwen2023relaxinghardware}. 
The total execution time for this operation is $3d + 2$ code cycles = 3 clocks (In the following, we ignore the small constant-time operations in the count of the lattice surgery clock). 
In certain cases, we do not need the patch to return to the initial position. 
Thus, the runtime decreases to 2 clocks.

\subsubsection{CNOT and CZ gate}
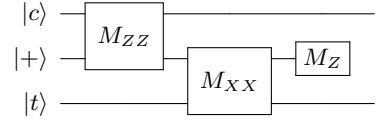
\begin{figure}
    \centering
    \mbox{
        \Qcircuit @C=1em @R=.7em {
        \lstick{\ket{c}} & \multigate{1}{M_{ZZ}} & \qw & \qw & \qw \\
        \lstick{\ket{+}} & \ghost{M_{ZZ}} & \multigate{1}{M_{XX}} & \gate{M_Z} \\
        \lstick{\ket{t}} & \qw  & \ghost{M_{XX}} & \qw & \qw 
        }
    }    
    \caption{Logical CNOT gate circuit using an ancilla qubit and $P\otimes P$ ($P = X, Z$) measurements. $\ket{c}$ and $\ket{t}$ indicate control and target qubits in the logical CNOT operation.}
    \label{fig:cnot_circuit}
\end{figure}

\begin{figure}
    \centering
    \includegraphics[width=80mm, clip]{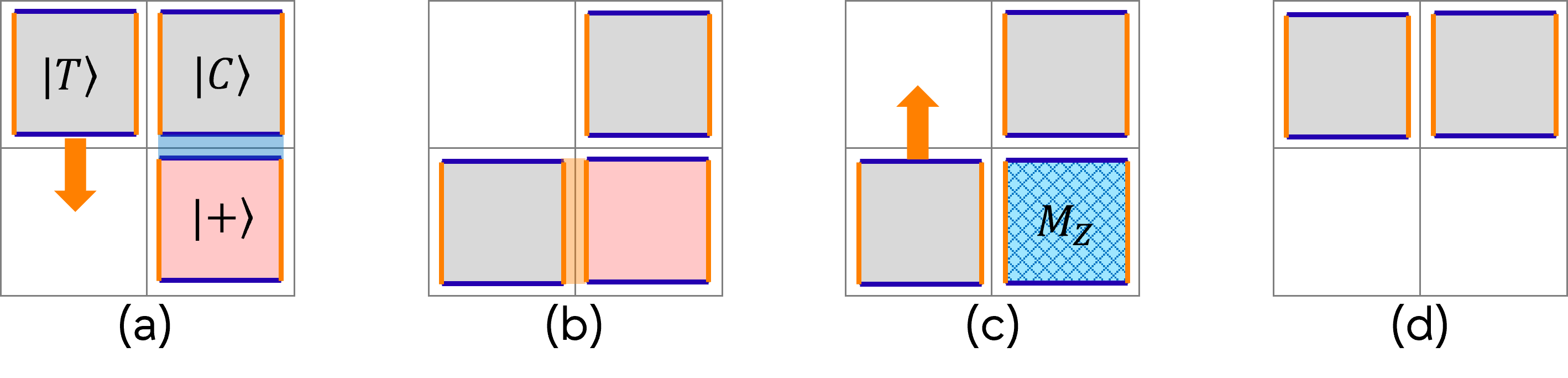}
    \caption{Logical CNOT gate using lattice surgery. 
    (a) An ancilla state is initialized to $\ket+$, and then the first $ZZ$ measurement is performed. 
    (b) $XX$ measurement between the ancilla state and $\ket T$. 
    (c) The destructive logical $Z$ measurement is performed on the ancilla state. 
    (d) Moving $\ket T$ to the initial position. }
    \label{fig:cnot_ls}
\end{figure}

The CNOT gate is decomposed into sequential Pauli measurements using an ancilla state initialized to $\ket{+}$ as shown in Fig.~\ref{fig:cnot_circuit}. 
The $P\otimes P$ ($P = Z, X$) measurements can be performed by the combination of patch merging and patch splitting. 
The lattice surgery operations are schematically shown in Fig.~\ref{fig:cnot_ls} The total execution time is $3d$ code cycles = 3 clocks (note that if we can ignore the final moving operation, it finishes with $2d$ code cycles = 2 clocks). 

The CZ gate is constructed using both the CNOT gate and the Hadamard gate. 
The CZ gate circuit and its lattice surgery implementation are depicted in Figs.~\ref{fig:cz_circuit} and \ref{fig:cz_ls}. 
The total execution time is 4 clocks.

\begin{figure}
    \centering
    \mbox{
        \Qcircuit @C=1em @R=.7em {
        \lstick{\ket{c}} & \multigate{1}{M_{ZZ}} & \qw & \qw & \qw & \qw \\
        \lstick{\ket{+}} & \ghost{M_{ZZ}} & \gate{H} & \multigate{1}{M_{ZZ}} & \gate{M_X} \\
        \lstick{\ket{t}} & \qw & \qw & \ghost{M_{ZZ}} & \qw & \qw 
        }
    }    
    \caption{Logical CZ gate circuit. This circuit can be obtained by combining the logical CNOT gate circuit shown in Fig.~\ref{fig:cnot_circuit} and the logical Hadamard gate. Note that $\ket{c}$ and $\ket{t}$ are interchangeable. }
    \label{fig:cz_circuit}
\end{figure}
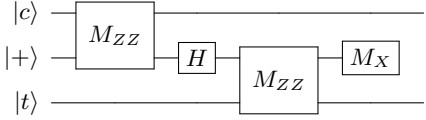
\begin{figure}
    \centering
    \includegraphics[width=80mm, clip]{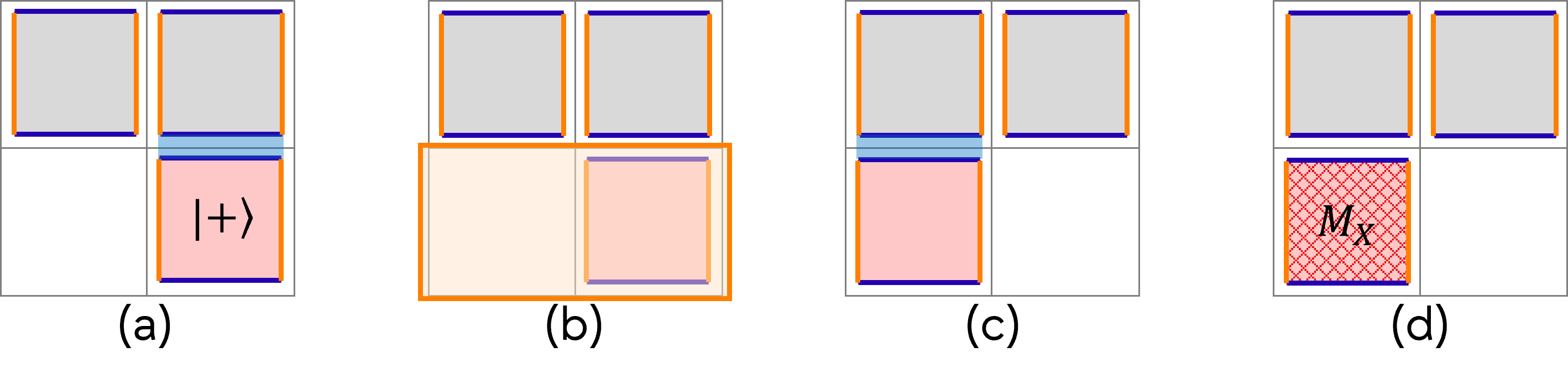}
    \caption{Logical CZ gate using lattice surgery.
    (a) An ancilla state is initialized to $\ket+$, and then the first $ZZ$ measurement is performed. 
    (b) After the $ZZ$ measurement, the transversal $H$ gate is applied on the ancilla state. Using the orange region, the patch rotation as well as patch moving is performed.   
    (c) The second $ZZ$ measurement is performed. 
    (d) The destructive logical $X$ measurement on the ancilla patch is performed.}
    \label{fig:cz_ls}
\end{figure}

\subsubsection{Multi-target CNOT and multi-target CZ gate} \label{appx:multicnotcz}
\begin{figure} 
    \centering
    \includegraphics[width=80mm, clip]{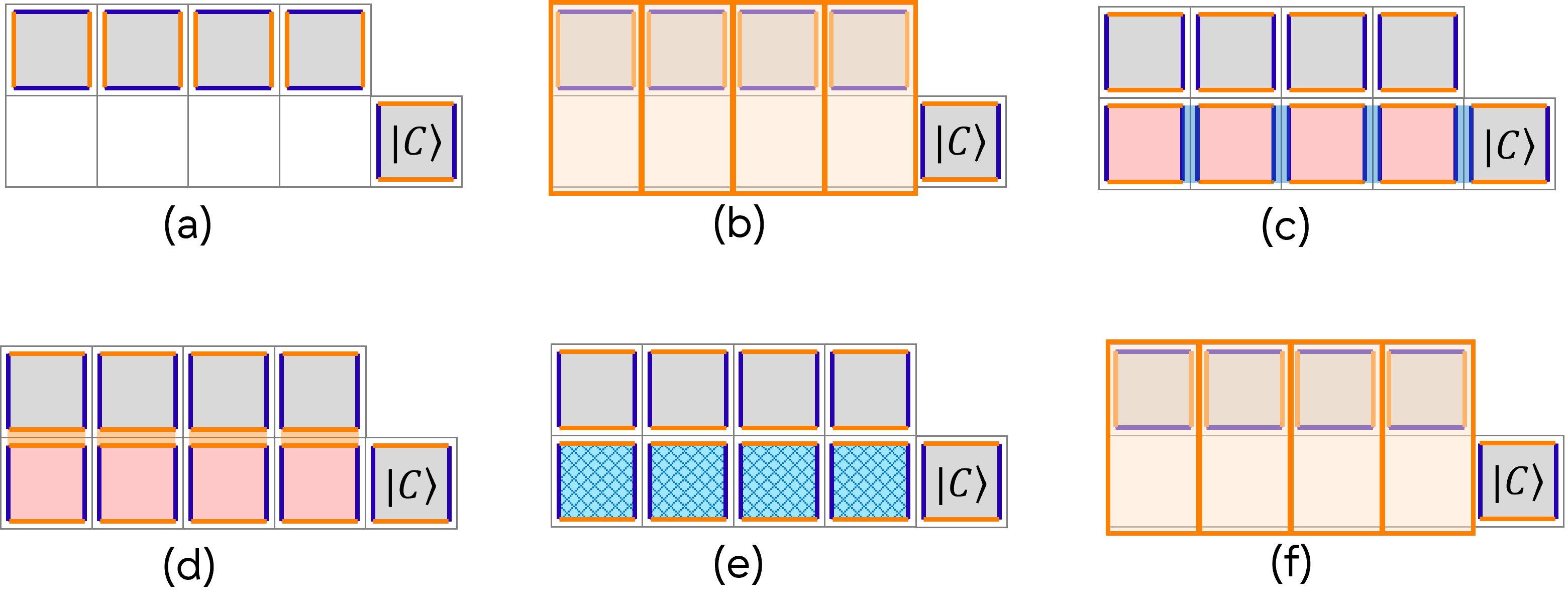}
    \caption{Multi-target logical CNOT operation using lattice surgery. 
    (a) Initial configuration with logical patches arranged as Fig.~\ref{fig:patch2nf2f}. We pick up the part of it in this figure. The lower-right patch is the control logical qubit, and the other four patches are the CNOT operation targets.
    (b) First, we rotate the target patches. 
    (c) Then, we prepare logical $\ket +$ in the ancilla regions (red patches) and perform logical $ZZ$ measurements. 
    These measurement results enable us to reproduce the parity between the control qubit and each ancilla patch. 
    (d) After the $ZZ$ measurements, we perform $XX$ measurements between the ancilla patches and the target patches. 
    (e) Then, the ancilla patches are measured on a $Z$ basis. Using the measurement results, appropriate Pauli corrections are applied (in practice it can be traced by software.)
    (f) The target patches are rotated again. 
    In total, these sequences take 8 clocks.
    }
    \label{fig:multitarg_cnot_ls}
\end{figure}

\begin{figure} 
    \centering
    \includegraphics[width=80mm, clip]{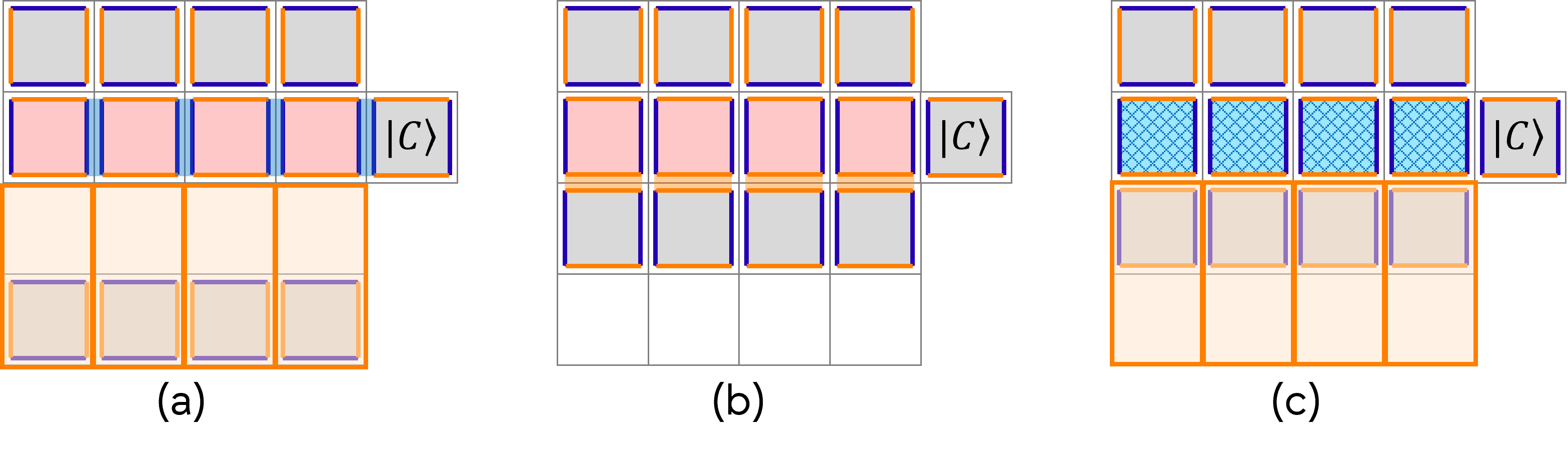}
    \caption{
    Runtime reduction in the multi-target CNOT gate using available ancilla region. 
    (a) Patch rotations and $ZZ$ measurements can be performed at the same time. 
    (b) $XX$ measurements.
    (c) Destructive $Z$ measurements and patch rotations. 
    The runtime decreases to 5 clocks in total. 
    }
    \label{fig:multitarg_cnot_ls_reduced}
\end{figure}

In this study, we also use the multi-target CNOT and CZ gates to realize the controlled time-evolution operator. 
Using lattice surgery, multi-target CNOT gates can be executed without decomposing it to the multiple single-target CNOT gates~\cite{Litinski2018latticesurgery}.
For a detailed example, see Fig.~\ref{fig:multitarg_cnot_ls} where the multi-target CNOT operation takes a total of 8 clocks.
Our compilation applies the multi-target CNOT only to a certain spin component (e.g. spin-down) as shown in Eq.(\ref{eq:K1}). 
By adopting the patch arrangement discussed in Sec.~\ref{sec:compile_ls}, we can reduce the runtime to 5 clocks, as shown in Fig.~\ref{fig:multitarg_cnot_ls_reduced}. 
This reduction assumes an initial division of spin-up and spin-down components into upper and lower rows, which introduces an additional patch moving overhead of 3 clocks at the beginning and end of the entire Trotter time evolution. This slightly differs from our discussion in Sec.~\ref{sec:compile_ls} (see Fig.~\ref{fig:init_Z_ZZ}).
Similarly to the cancellation of the multi-target CNOT gate, these patch-moving operations also cancel with each other between the Trotter steps.

The multi-target CZ gate operates similarly but includes 
additional $H$ operations on target patches. 
Fortunately, the $H$ operation is immideate (counted as 0 clock) and automatically rotates the patch,  
eliminating the need for operations (b) and (f) in Fig.~\ref{fig:multitarg_cnot_ls}, and the runtime becomes just 2 clocks.
In the main text, since we need to perform multi-target CZ operations on both up and down spin components, the total runtime is $2 \times 2 = 4$ clocks.


\subsubsection{S gate}
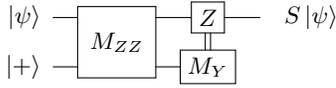
\begin{figure}[tbp]
    \centering
    \mbox{
        \Qcircuit @C=1em @R=.7em {
        \lstick{\ket{\psi}} & \multigate{1}{M_{ZZ}} & \gate{Z}        & \qw & \push{S \ket{\psi}} \\
        \lstick{\ket{+}}    & \ghost{M_{ZZ}}        & \gate{M_Y} \cwx &
        }
        }    
    \caption{Logical $S$ gate circuit using the in-place $Y$ measurement. $M_{ZZ}$ is performed using the standard merging-splitting operation. The destructive $Y$ measurement $M_Y$ is performed by the technique proposed in Ref.~\cite{Gidney2024inplaceaccessto}. 
    Since $M_{ZZ}$ and $M_Y$ takes $d$ and $d/2$ code cycles for each, this circuit takes $d+d/2$ code cycles = 1.5 clocks in total.}
    \label{fig:sgate_circuit}
\end{figure}

\begin{figure}
    \centering
    \includegraphics[width=50mm, clip]{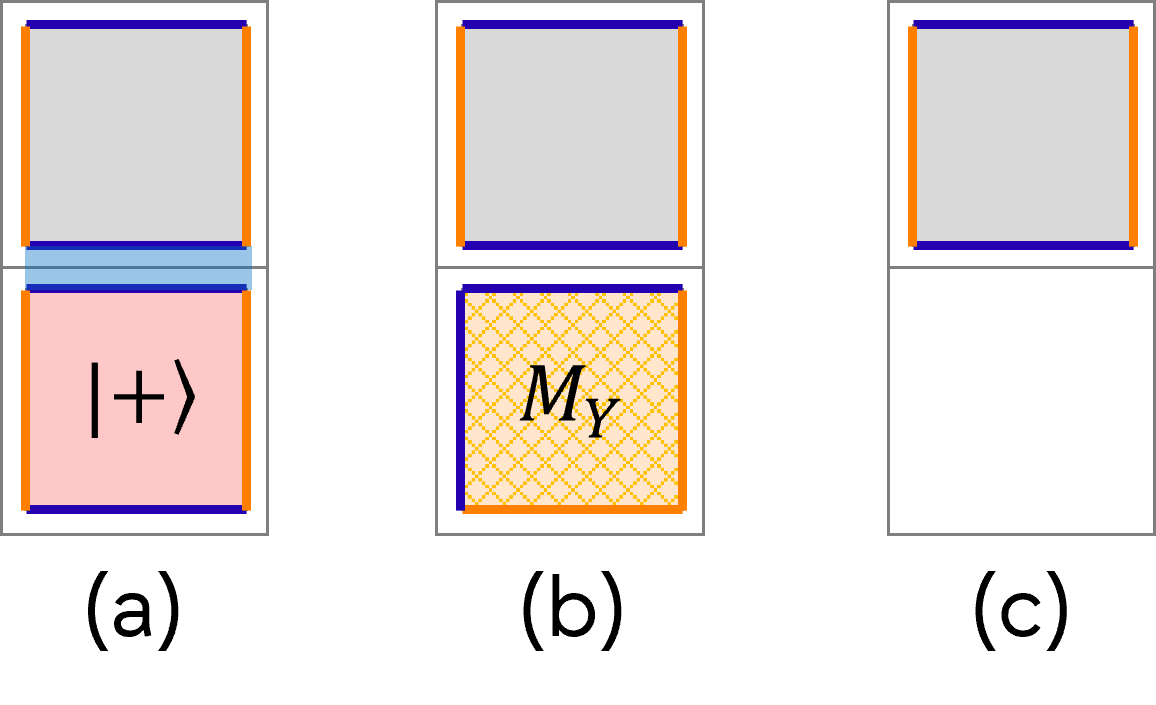}
    \caption{Lattice surgery operations in logical $S$ gate (circuit of Fig.~\ref{fig:sgate_circuit}). Only one additional patch is required.
    (a) The logical $ZZ$ measurement between the target patch and the ancilla patch is performed. 
    (b) The in-place $Y$ basis measurement on the ancilla patch is performed. This operation can be performed within $ \lfloor d/2 \rfloor$ clocks~\cite{Gidney2024inplaceaccessto}.
    (c) Final configuration. 
    }
    \label{fig:s_ls}
\end{figure}

The implementation of the $S$ gate on the rotated planar surface code is notably complex since the surface code is not self-dual.
Traditionally, the $S$ gate has been realized through the gate teleportation using a special resource state (the eigenstate of the Pauli $Y$ operator)~\cite{PhysRevA.86.032324} or by performing a logical $Y$ basis measurement using a rectangular-shaped patch~\cite{Litinski2019gameofsurfacecodes}. 
Recently, a new method has been proposed for 
the fault-tolerant $Y$ basis measurement within a single rotated planar surface code patch~\cite{Gidney2024inplaceaccessto}. 
This $Y$ basis measurement enables us to perform $S$ gate with minimal space overhead (only two patches) as shown in the quantum circuit in Fig.~\ref{fig:sgate_circuit}. 
We employ this efficient method for the  $S$ gate implementation. 
The total execution time is $d+d/2$ code cycles = 1.5 clocks, as schematcally shown in Fig.~\ref{fig:s_ls}. 

\section{Small-angle analog rotation gates} \label{appx:rotationgate}
In this section, we discuss the implementation of the analog rotation gate. 
We first introduce the RUS protocol and then discuss the ancilla state injection protocol~\cite{Toshio2024}, which is suitable for the small-angle rotation gates.  

Instead of using Clifford + $T$ decomposition, the 
STAR architecture directly performs the analog rotation gate 
by consuming a resource state, 
$\ket{m_\theta} \equiv R_Z(\theta) \ket + = \frac{1}{\sqrt{2}}(e^{-i \theta} \ket 0 + e^{+i \theta} \ket 1)$, 
where the angle $\theta$ is assumed to be small. 
As shown in Fig.~\ref{fig:rotation_circuit}, the circuit for the analog rotation gate produces different output states depending on the measurement result in the $ZZ$ basis. If the measured value is $+1$ ($-1$), the output state is $R_{Z}(\theta) \ket{\psi}$ ($R_{Z}(-\theta) \ket{\psi}$). 
If we obtain $R_{Z}(-\theta) \ket{\psi}$, we perform the same circuit with updated angle $\theta' = 2\theta$ to compensate for the wrong rotation angle $-\theta$. 
The second circuit run is also probabilistic, so we repeat this procedure until obtaining the desired state $R_{Z}(\theta) \ket{\psi}$. 
The protocol is called the RUS protocol~\cite{Cody_Jones_2012,reiher2017elucidating,PRXQuantum.5.010337}. 
Each run of the circuit in Fig.~\ref{fig:rotation_circuit} during the RUS protocol is called an ``RUS trial,'' and an instance of the protocol (a single rotation gate) is called an ``RUS process.''
On average, a single RUS process is completed in two trials. 

\begin{figure}
    \centering
    \includegraphics[width=70mm, clip]{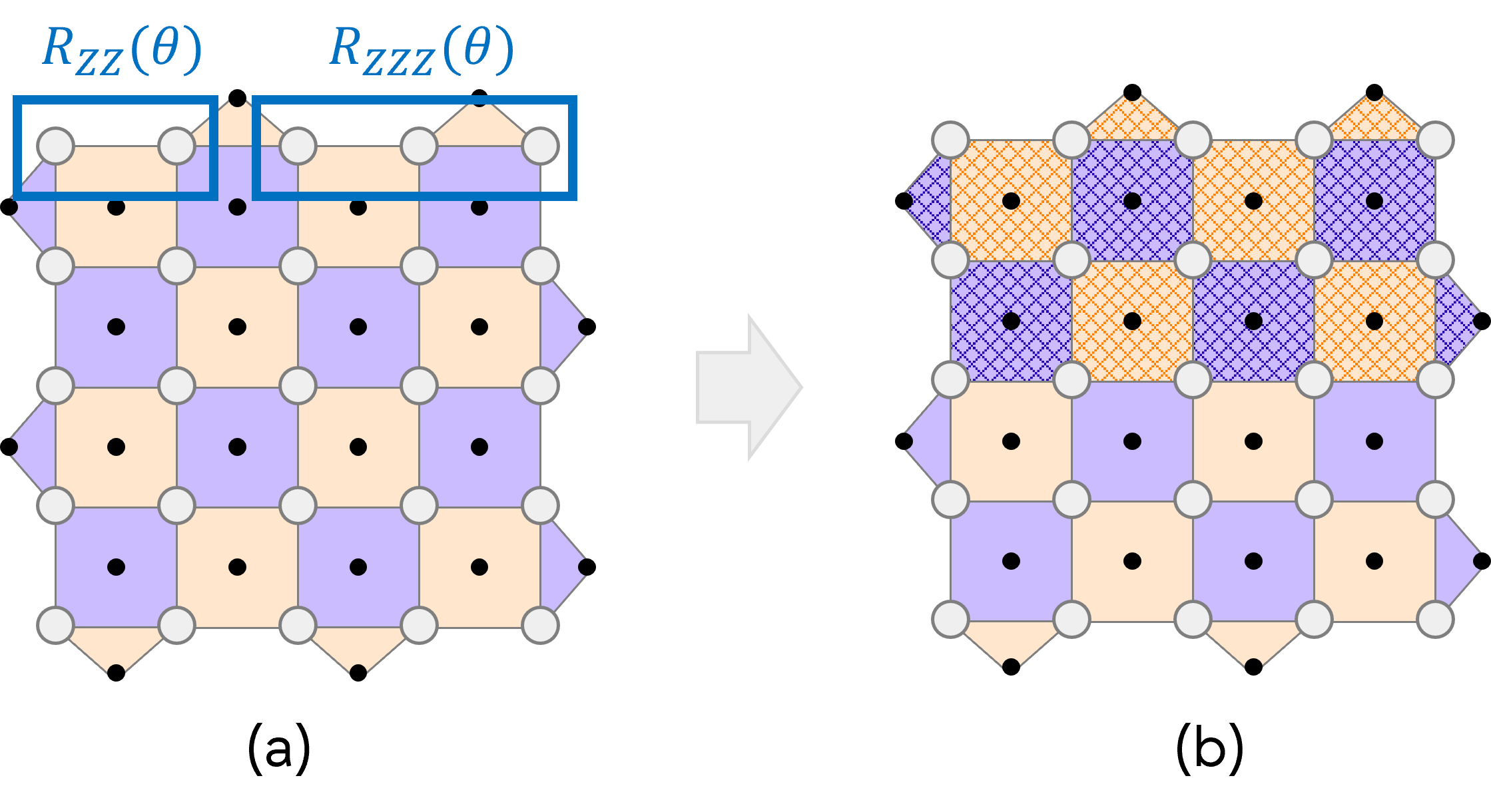}
    \caption{
    Ancilla state injection protocol in STAR architecture. 
    (a) The ancilla state is initialized in $\ket{+}_L$ first, then multiple rotation gates are acted along the logical $Z_L$ operator. Multi-Z rotation gates can be performed using CNOT gates and Z rotation gates.
    (b) After acting the rotation gates, we perform a post-selection on the hatched stabilizers to remove critical errors that bring leading logical errors. Errors not detected in the post-selection can be corrected in the subsequent quantum error correction procedure safely. 
    }
    \label{fig:injection}
\end{figure}

Although all of the logical measurements in Fig.~\ref{fig:rotation_circuit} are fault-tolerant via the  lattice surgery, 
the injection of the ancilla state $\ket{m_\theta}$ cannot be fault-tolerant owing to the analog nature of rotation angle. 
Therefore, the precision of the injected ancilla state directly affects the precision of the rotation gate. 
Here, we briefly introduce how to prepare the high-quality ancilla state for small-angle rotation in the STAR architecture~\cite{Toshio2024} (For a more in-depth discussion, refer to the original paper~\cite{Toshio2024}). 
The injection protocol sequence is depicted in Fig.~\ref{fig:injection}.
To inject $\ket{m_{\theta^*}}$, where $\theta^*$ is a small target angle, 
we begin with a patch initialized in the $\ket{+}$ state, which can be prepared fault-tolerantly using standard methods. 
We define a subset of physical qubits that support the logical $Z_L$ operator as $Q_{Z_L}$, such that  $\prod_{j \in Q_{Z_L}} Z_j = Z_L$. 
To construct the desired ancilla state, we perform the following multiple rotation gates on the logical $\ket{+}$ state: 
\begin{equation}
    \prod_{i=0}^{k-1} R_{Z_{Q_i}}(\theta) = \prod_{i=0}^{k-1} \left[ \cos \theta I_{Q_i} + i \sin \theta Z_{Q_i} \right],  
\end{equation}
where $Q_i$ ($i = 0, \dots k-1$) is $i$-th subset of physical qubits, which satisfies $\bigcup_i Q_i = Q_{Z_L}$. 
$I_{Q_i}$ and $Z_{Q_i}$ denote the identity and multi-qubit $Z$ operator acting on the subset $Q_i$, respectively. 
An example is shown in Fig.~\ref{fig:injection} (a), which depicts $k=2$ rotation gates. 
The resultant state after some Clifford corrections is a superposition of orthogonal states as follows: 
\begin{equation} \label{eq:ver2_state}
    \prod_{i=0}^{k-1} R_{Z_{Q_i}}(\theta) \ket{+} = \sqrt{p_{\rm ideal}} \ket{m_{\theta^*}} + (\text{orthogonal states}), 
\end{equation}
where the angles $\theta$ and $\theta^*$ satisfy 
\begin{equation}
    \theta^* = \sin^{-1} \left( \frac{1}{\sqrt{p_{\rm ideal}}} \sin^k (\theta) \right) \approx \theta^k + {\mathcal O}(\theta^{k+2}).
\end{equation}
$p_{\rm ideal}$ and ``orthogonal states'' in Eq.(\ref{eq:ver2_state}) are needed to explain.  
States other than the target state in Eq.(\ref{eq:ver2_state}) are in orthogonal stabilizer spaces owing to additional $Z$ operators acting on these states. 
Therefore, the syndrome measurement projects the superposed state into one of these states. 
The $p_{\rm ideal}$ value denotes the probability of obtaining the correct state through the syndrome measurement, 
\begin{equation}
    p_{\rm ideal}(\theta, k) = \sin^{2k} \theta + \cos^{2k} \theta \approx 1 - k \theta^2 + {\mathcal O}(\theta^4).  
\end{equation}
As clearly seen, $p_{\rm ideal}$ is nearly identity when the rotation angle is very small. 
Of course, since there are quantum noises in practice, we need to remove noisy states as many as possible. 
To address this, we measure the error syndrome twice and perform postselection. 
The stabilizer operators checked in postselection are shown by the hatched pattern in Fig.~\ref{fig:injection} (b). 
Errors outside the postselected region can be corrected by subsequent QEC procedures and do not lead to leading-order logical errors. 
This small-size postselection enables us to generate a high-quality ancilla state with a high success rate. 
Our state injection protocol, which can be performed within a single patch, 
eliminates the need for an ``ancilla state factory.'' 
Thanks to the compactness of our rotation gate, we can perform multiple rotation gates in parallel without additional space overhead 
provided that there are enough free patches enough for the state injection. 
This ``easy-to-parallelize'' feature of the STAR architecture contrasts significantly with existing FTQC architectures, where parallel $T$ gates require many more physical qubits to parallelize the magic state distillation. 

\section{Derivation of Eq.(\ref{eq:aveRUS})} \label{appx:averageRUSderiv}
In this section, we outlines the derivation of Eq.(\ref{eq:aveRUS}). 
Let us consider the case that involves performing $M$ independent rotation gates in parallel.
To estimate the average RUS trial number needed to complete all $M$ processes, 
we first calculate the probability that the parallel RUS execution finishes at the $K$-th RUS trial, $P_{K}^{M}$. 
This scenario can be divided into $M$ independent cases further: $M-1$ processes finish before the $K$-th trial, then the remaining process finishes on the $K$-th trial; 
$M-2$ processes finish before the $K$-th trial, and the remaining two processes finish on the $K$-th trial; 
and so on. 
Since the single RUS trial finishes with a probability of $1/2$, we can calculate the probability of these $M$ cases. 
For example, the probability of the first case is the following, 
\begin{eqnarray}
    p^{M}_{\text{one process finishes at} K} &=& M \left( \sum_{n=1}^{K-1} \left(\frac{1}{2} \right)^n \right)^{M-1} \cdot \left(\frac{1}{2} \right)^K \nonumber \\ 
    &=& M A^{M-1} B,
\end{eqnarray}
where we define $A \equiv \sum_{n=1}^{K-1}(1/2)^n$ and $B \equiv (1/2)^K$. 
Factor $M$ represents the number of choices for which process finishes on the $K$-th trial. 
$A^{M-1}$ is a sum of probability for all possible patterns that $M-1$ RUS processes finish before the $K$-th RUS trial. 
Other cases are calculated similarly. 
As a result, the sum of probabilities of all cases, which is $P_K^M$, is given as,
\begin{equation} \label{eq:probkm}
    \begin{array}{ll}
    P_K^M &= MA^{M-1}B + {}_M C_2 A^{M-2} B^2 + \cdots + B^M \\
    &= (A+B)^M - A^M, 
    \end{array}
\end{equation}
where $_M C_L$ is the binomial coefficients, 
and we use the binomial expansion $(a+b)^M = a^M + M a^{M-1}b + \cdots + b^M$ at the second equality. 
Therefore, the average number of RUS trials needed to complete all $M$ RUS processes is provided as follows, 
\begin{equation} 
    \begin{array}{ll}
    \langle K \rangle_M &= \sum_K K P_K^M \\
    &= \sum_K K \left[ (1-2^{-K})^M - (1-2^{-K+1})^M \right], 
    \end{array}
\end{equation}
where we use $\sum_{n=1}^{L} (1/2)^n = 1 - 2^{-L}$. 

\section{Controlled time evolution operator} \label{appx:ctrlrot}
\begin{figure}
    \centering
    \includegraphics[width=80mm, clip]{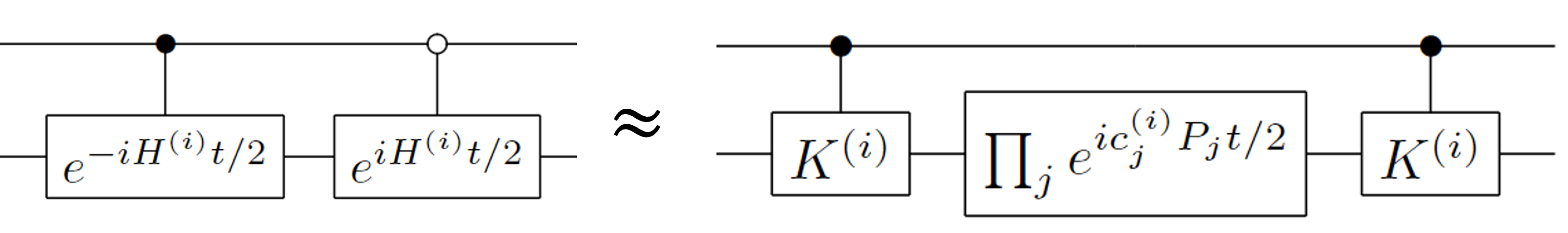}
    \caption{Example of the controlled time evolution by a combination of the controlled multi-qubit Pauli operator and control-free time evolution.}
    \label{fig:ctrl_time_evolution}
\end{figure}

The controlled time evolution operator appears in several quantum algorithms. 
A typical example is the Hadamard test circuit shown in Fig.~\ref{fig:hadamardtest}. 
The naive approach to implementing controlled time evolution involves replacing each rotation gate $e^{-i c_j P_j \tau}$ in the Trotter step to the controlled counterparts. 
However, this naive replacement introduces an overhead proportional to the number of interaction terms, which becomes impractical for large system size. 
Fortunately, for certain Hamiltonians, efficient implementation of controlled time evolution is possible with a reasonable overhead~\cite{PRXQuantum.3.040305}. 
Let us consider a Hamiltonian consisting of a certain set of sub-Hamiltonians, 
\begin{equation}
    H = \sum_l H^{(l)},\quad H^{(l)} = \sum_j c^{(l)}_j P_j,
\end{equation}
where $P_j$ is a multi-qubit Pauli operator and $c^{(l)}_j$ is a coefficient of the sub-Hamiltonian $H^{(l)}$. 
Let $K^{(l)}$ be a certain multi-qubit Pauli operator that anti-commutes with the sub-Hamiltonian $H^{(l)}$, $K^{(l)} H^{(l)} K^{(l)} = -H^{(l)}$. 
This relationship leads to $K^{(l)} e^{+i H^{(l)} t} K^{(l)} = e^{-i H^{(l)} t}$. 
In this case, we can easily construct a controlled time evolution operator 
by combining the control-free time evolution and controlled-$K^{(l)}$ operations, as shown in Fig.~\ref{fig:ctrl_time_evolution}. 
This method is well-suited for our purpose since 
it allows us to leverage the compilation results directly.

\begin{figure}
    \centering
    \includegraphics[width=30mm, clip]{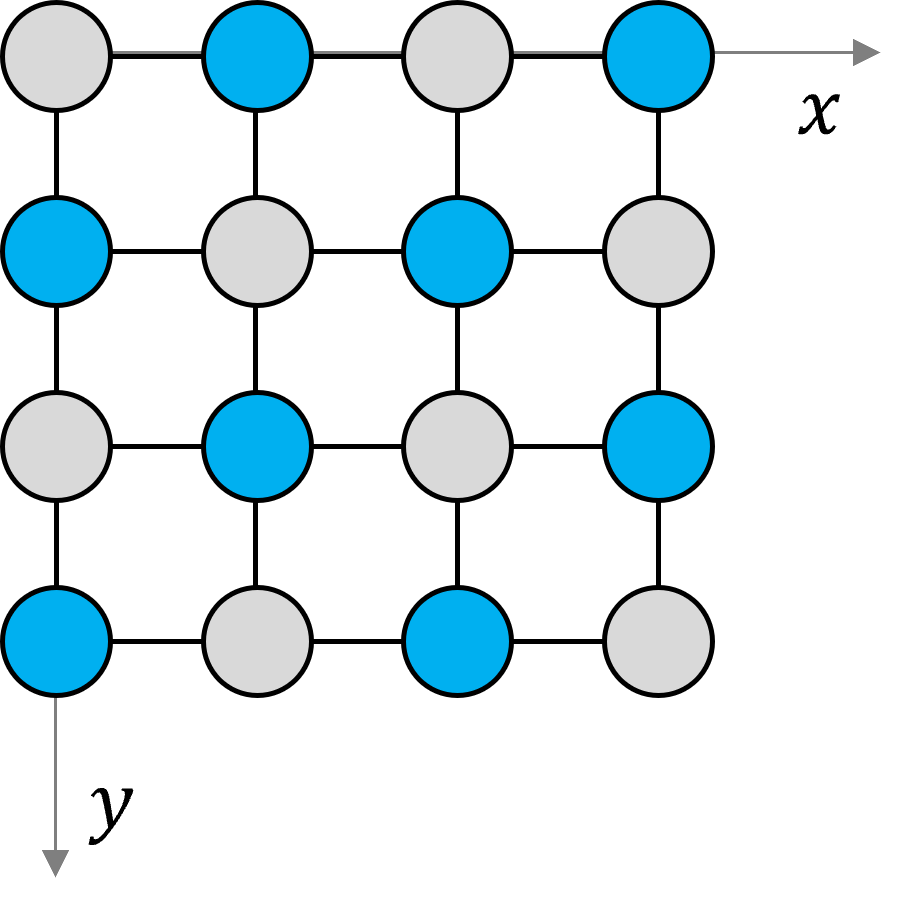}
    \caption{Targets of the CZ operations on the 2D lattice ($4 \times 4$). 
    The blue nodes are the target of the CZ operations. 
    Here we only show up spin components.
    The upper-left node is the origin of the coordinates, $(x,y) = (0,0)$.}
    \label{fig:CZ_targets}
\end{figure}

For the 2D Hubbard model, we can choose the sub-Hamiltonian as follows: 
\begin{equation}
    H = H^{(0)} + H^{(1)},
\end{equation}
\begin{eqnarray}
    H^{(0)} &=& -\frac{t}{2} \sum_{\langle i,j \rangle, \sigma} (X_{i,\sigma} X_{j,\sigma} + Y_{i,\sigma} Y_{j,\sigma}) Z^{\leftrightarrow}_{i,j,\sigma}, \\
    H^{(1)} &=& \frac{U}{4} \sum_{i} Z_{i,\uparrow} Z_{i,\downarrow}.
\end{eqnarray}
In this choice, $K^{(l)}$ ($l = 0,1$) can be chosen as 
\begin{eqnarray}
    K^{(0)} &=& \prod_{k \in V^{(0)}, \sigma} Z_{k, \sigma}, \\
    K^{(1)} &=& \prod_{k \in V^{(1)}} X_{k, \downarrow}. \label{eq:K1} 
\end{eqnarray}
where $V^{(l)}$ ($l = 0,1$) are defined as 
\begin{eqnarray}
    V^{(0)} &=& \{k = (x,y) \mid x+y = 1 \mod 2 \}, \\
    V^{(1)} &=& \{ \forall k = (x,y) \}, 
\end{eqnarray}
where we introduce the 2D coordinate of the Hubbard model lattice for the first time, $k = (x, y)$ ($x=0,\dots,N-1, y=0,\dots,N-1$). 
We show the example of $V^{(0)}$ on the $4 \times 4$ lattice in Fig.~\ref{fig:CZ_targets}. 
Furthermore, since the control qubit is common for all CNOT and CZ operations, 
multiple operations can be combined into a single multi-target operation.
For example, we can merge multiple CNOT gates with the same control qubit into a single multi-target CNOT gate using an appropriate lattice surgery operation~\cite{Litinski2018latticesurgery}. 
See Appendix~\ref{appx:multicnotcz} for more details. 
The multi-target CNOT gate has a constant time overhead, allowing us to minimize additional time overhead for controlled time evolution. 
{
Typically, four multi-target CNOT layers and four multi-target CZ layers are required for each Trotter step to construct the controlled time evolution operator. 
However, two multi-target CZ layers in the middle of the Trotter formula cancel each other out. 
In addition, the first and last multi-target CNOT layers cancel with those from the previous and next Trotter steps. 
By considering the lattice surgery clock discussed in Appendix~\ref{appx:multicnotcz}, 
the total overhead of the controlled time evolution operator, consisting of $T$ Trotter steps, can be estimated as follows, 
\begin{equation}
    T_{\rm ctrl} = 2\cdot 5 + 2\cdot 3 + T (2 \cdot 5 + 2 \cdot 4) = 16 + 18T, 
\end{equation}
where the first two terms correspond to the overhead of multi-target CNOT and patch moving operations at the beginning and the end of the Trotter time evolution. 
}

\section{QCELS algorithm} \label{appx:qcels}
In the following, we first introduce the main idea of the QCELS algorithm in a simple set-up and then cite Algorithm 1 and Theorem 2 in Ref.~\cite{PRXQuantum.4.020331}, which we need for the resource estimation. 

The QCELS algorithm relies on a very simple idea to extract the eigenvalues. 
Let us consider we have the time-series data of the expectation value, $Z_n = \bra{\psi} e^{-i t_n H} \ket{\psi}$ ($t_n = n \tau$ for $n = 0,1, \dots, N-1$). 
If the initial state $\ket{\psi}$ overlaps the ground state $\ket{E_0}$ largely, $p_0 \approx 1$, 
the expectation value approximately behaves as $Z_n \approx p_0 e^{-i E_0 t_n}$.  
Therefore we can extract the energy eigenvalue by solving the following non-linear least-squares problem: 
\begin{eqnarray}
    \left( r^*, \theta^* \right) &=& {\rm argmin}_{r, \theta} L(r,\theta), \\
    L(r, \theta) &=& \frac{1}{N} \sum_{n=0}^{N-1} |Z_n - r e^{-i t_n \theta}|^2. \label{eq:loss_func}
\end{eqnarray}
This optimization problem can be solved classically. 
The optimal $\theta^*$ provides an approximation of the energy eigenvalue $E_0$. 
This optimization subroutine is why the algorithm is named ``quantum complex exponential least squares'' (QCELS). 

The basic concept of the QCELS algorithm is straightforward, but to achieve Heisenberg-limited scaling, the actual implementation is more complex.
The whole algorithm, called ``multi-level QCELS,'' is cited in Algorithm 1 to make this manuscript self-contained. 
In multi-level QCELS, we start with a small time evolution interval $\tau_1$ and estimate the eigenvalue using the time-series data $Z^1_n = \bra{\psi} e^{-i n \tau_1 H} \ket{\psi}$ ($n = 0,1, \dots, N-1$). 
Small $\tau_1$ results in a rough estimate of $E_0$; thus, we refine this estimation by increasing the time interval $\tau_1$ to $\tau_j = 2^{j-1} \tau_1$ ($j = 1,\dots, J$), where $J \propto \log_2(\delta/N\epsilon)$, $\delta$ is a small constant (their definitions are provided in Theorem 2 cited below), and $\epsilon$ is the target precision of the estimation of $E_0$. 
Performing the estimation of $E_0$ $J$ times with increasing $\tau_j$, we achieve an estimation of $E_0$ with precision $\epsilon$. 
\begin{figure}
\begin{algorithm}[H] \label{alg:alg1}
    \caption{multi-level QCELS algorithm in Ref.~\cite{PRXQuantum.4.020331}, p.8}
    \begin{algorithmic}[1] 
    \State \textbf{Preparation:} Number of data pairs: $N$; number of samples: $N_s$; number of iterations: $J$; sequence of time steps: $\{\tau_j\}^{J}_{j=1}$; Quantum oracle: $\left\{\exp(-i\tau_j H)\right\}^J_{j=1
  }$;
    \State \textbf{Running:}
    \State $\lambda_{\min}\gets-\pi$; $\lambda_{\max}\gets\pi$; \Comment{$[\lambda_{\min},\lambda_{\max}]$ contains $\lambda_0$}
    \State $j\gets 1$;
    \For{$j=1,\ldots,J$}
    \State Generate the data set in $\mathcal D_H = \{ (n\tau_j, Z_n) \}_{n=0}^{N-1}$ using the circuit in Fig.~\ref{fig:hadamardtest} with $t_n=n\tau_j$.
  
    \State Define loss function $L(r,\theta)$ according to Eq.(\ref{eq:loss_func}).
    \State Minimizing the loss function.
  \[
      (r^*_j,\theta^*_j)\gets \underset{r\in\mathbb{C},\theta\in[-\lambda_{\min},\lambda_{\max}]} {\rm argmin} L(r,\theta)\,,
  \] 
  \State $\lambda_{\min}\gets\theta^*_j-\frac{\pi}{2\tau_j}$;  $\lambda_{\max}\gets\theta^*_j+\frac{\pi}{2\tau_j}$ \Comment{Shrink the search interval by $1/2$}
  \EndFor
  \State \textbf{Output:} $\theta^*$
  \end{algorithmic}
\end{algorithm}
\end{figure}
The computational complexity of Algorithm 1 is stated in Theorem 2 of Ref.~\cite{PRXQuantum.4.020331}, 
which introduces key parameters for resource estimation. 
\setcounter{thm}{1}
\begin{thm}[Ref.\cite{PRXQuantum.4.020331}, p.9]
    Let $\theta^*$ be the output of Algorithm \ref{alg:alg1}.
    Given $p_0>0.71$, $0<\eta<1/2$, $0<\epsilon<1/2$, we can choose $\delta$ according to $\delta = \Theta (\sqrt{1-p_0})$,  
    \begin{equation}
    \begin{array}{ll}
        J=\left\lceil\log_2(1/\epsilon)\right\rceil+1, \\ 
        \tau_j=2^{j-1-\left\lceil\log_2(1/\epsilon)\right\rceil}\frac{\delta}{N\epsilon}, \quad \forall 1\leq j\leq J\,.
    \end{array}
    \label{eqn:tauj_choice}
    \end{equation}
    Choose $NN_s= \widetilde{\Theta}\left(\delta^{-(2+o(1))}\right)$. Then 
    \begin{equation} \label{eq:t_max}
    T_{\max}=N\tau_J=\frac{\delta}{\epsilon}, 
    \end{equation}
    \begin{equation} \label{eq:t_total}
    T_{\mathrm{total}}=\sum^J_{j=1}N(N-1)N_s\tau_j/2=\widetilde{\Theta}\left(\delta^{-(1+o(1))}\epsilon^{-1}\right)\,
    \end{equation}
    and
    \[
    \mathbb{P}\left(\left|(\theta^*-\lambda_0)\bmod [-\pi,\pi)\right|<\epsilon\right)\geq 1-\eta\,.
    \]
\end{thm}
The resource estimation in this study relies on the complexity of Theorem 2 cited above. 

\section{Runtime of the second-order Trotter decomposition using the serial compilation strategy} \label{appx:serialcompile}
In this appendix, we derive the runtime for the second-order Trotter decomposition using the serial compilation strategy~\cite{Litinski2019gameofsurfacecodes}. 
To begin with, we count
how the number of rotations required and how to perform them. 
We consider the same Hamiltonian as in the main text, namely Eq.(\ref{eq:2Dhubbard_final}). 
In this Hamiltonian, there are $V$ $ZZ$ rotation terms, $4(V-M)$ $XZ\dots ZX$ rotation terms, and $4(V-M)$ $YZ \dots ZY$ rotation terms. 
The second-order Trotter decomposition doubles these numbers except for the two rotation layers in the middle, which can be merged into a single rotation layer. 
For the purpose of this estimation, we place the $ZZ$ rotation gate layer in the middle of the decomposition. 
Therefore, the runtime of the second-order Trotter decomposition using the serial compilation is given as 
\begin{equation} \label{eq:serial_runtime_per_trott}
    VT_{ZZ} + 2\cdot 4(V-M)T_{XZ\dots ZX} + 2\cdot 4(V-M)T_{YZ\dots ZY}, 
\end{equation}
where $T_{P}$ is an average runtime of the Pauli $P$ rotation gate.

Next, we discuss the runtime estimation of each rotation gate. 
In the context of estimating the RUS overhead, we assume that the injection overhead can be hidden by separating the injection patches from the calculation block, as discussed in Ref.~\cite{PRXQuantum.5.010337}. 
Moreover, since each rotation gate is performed sequentially, the average trial of the RUS protocol is 2. Therefore, assuming Pauli measurements requires 1 clock, the average runtime of each RUS protocol is 2 clocks. 
For the $ZZ$ rotation gates, no basis change operations are needed, allowing them to be performed in $T_{ZZ} = 2$ clocks. 
The $XZ \dots ZX$ rotation gates need patch rotation at the beginning and the end, resulting in an average runtime of $T_{XZ\dots ZX} = 3\times 2 + 2 = 8$ clocks. 
Similarly, the $YZ \dots ZY$ rotation gates need patch rotation, and the $S$ gates at the beginning and the end, leading to an average runtime of $T_{YZ\dots ZY} = 3\times 2 + 1.5\times 2 + 2 = 11$ clocks on average. 
Therefore, by inserting these values in Eq.(\ref{eq:serial_runtime_per_trott}), we obtain the following, 
\begin{equation} 
    V\cdot 2 + 2\cdot 4(V-M) \cdot 8 + 2\cdot 4(V-M)\cdot 11 = 154V - 152M. 
\end{equation}
The comparison between the serial and parallel compilation results is displayed in Fig.~\ref{fig:vs_serial}.  

\bibliography{bib.bib}

\end{document}